# A self-organized compression network arrests epithelial proliferation

Liav Daraf[1], Yael Lavi[1], Areej Saleem[1], Daniel Sevilla Sanchez[2], Yuri Feldman[1], Lior Atia[1]*

[1]Department of Mechanical Engineering, Ben-Gurion University of the Negev, Beer-Sheva 8410501, Israel;
[2]Ilse Katz Institute for Nanoscale Science & Technology, Ben-Gurion University of the Negev, Beer-Sheva 8410501, Israel;
*Corresponding author. Email: atialior@bgu.ac.il

## Abstract

As epithelial development or wound closure approaches completion, cell proliferation progressively slows via contact inhibition of proliferation (CIP) - a mechanism understood as being strictly local. Here we report the discovery of inhibition of proliferation through an unanticipated mechanism that is non-local. As a confluent epithelial layer becomes progressively more jammed, two interpenetrating networks emerge: islands of mechanically compressed non-cycling cells percolating within an ocean of mechanically tensed cycling cells. The evolution of the compression network was found to be susceptible to both specific molecular stimulus and to injury-induced unjamming. Yet, in all circumstances, the size of compressed islands followed a power-law distribution that was well-captured by preferential network theory. Together, these findings demonstrate the existence of a network-based inhibition of proliferation (NIP) that is self-organizing and poised in proximity to criticality.

## Introduction

Epithelial cells line hollow organs and cavities, and participate in dynamic processes such as morphogenesis and wound healing. As these processes approach completion, the tissue gradually transitions to a homeostatic state in which cell divisions are suppressed enough to just balance cell apoptosis and extrusion[1]. Contributing to this suppression are various intracellular signaling pathways mediated by cell surface receptors, which together define what is known as contact inhibition of proliferation (CIP)[2-9]. By this very definition, CIP is understood as being a process that is strictly local. By contrast, here we report the discovery of an inhibitory process that is non-local. By spatially mapping the mechanical state of each cell —compressed vs. tensed— we show that inhibitory activity initiates in single compressed cells and then propagates to neighboring compressed cells. This process gradually forms a multiscale compression network that grows as the epithelial layer becomes less migratory and more jammed, and retreats as the layer becomes more migratory and less jammed[10-28].

## Two mechanical phenotypes of cells emerge with jamming

To examine spatial variations in the mechanical state of each cell in an epithelial layer, we began by exploring cell responses when adhesive constraints are alleviated and the cell therefore morphs towards its stress-free shape. Madin-Darby canine kidney (MDCK) cells were plated on a soft polyacrylamide gel coated with collagen to model a basal lamina. Inspired by earlier studies[29-33], we developed a restrained trypsinization assay to break all cell-cell and cell-substrate bonds (Fig. 1a-c; Movie S1). We then tracked and analyzed the temporal response of all cells in each examined field of view in terms of their two-dimensional (2D) and three-dimensional (3D) shapes. Since conventional thinking portrays the epithelium in a dominant state of tension, we expected all cells to contract post-deadhesion. Data showed not only a mechanical phenotype in which cells contracted as expected, but also, to our surprise, cells that expanded (Fig. 1d). Contracting cells tended to be larger in initial area, while expanding cells tended to be smaller with it. Yet, 2D cell area alone did not account for the expanding cell phenotype, as seen with vast numbers of small (pre-deadhesion) cells that did not expand (post-deadhesion) (Fig. 1d,e).

To explore 3D features, we used confocal microscopy on fixed and phalloidin-stained monolayers, either with no deadhesion (Fig. 1f, top) and post-deadhesion (Fig. 1f, bottom). Data revealed a transition from column-like to sphere-like geometry (Fig. 1g). From these observations (Extended Data Fig.1) we estimated changes in cell volume and surface area (Fig. 1h–i). Both expanding and contracting cells exhibited an increase in volume, which was correlated with the extent of 2D area change (Fig. 1h). However, the surface area increased only in expanding cells and decreased in contracting cells (Fig. 1i). This behavior is consistent with previous studies[32,33], and implies that surface area changes reflect membrane unfolding in expanding cells, and folding in contracting cells. We thus concluded that 3D features described similar general behavior of the post-trypsinization changes as in 2D (Fig. 1d).

With respect to temporal features during the deadhesion (arrow in Fig. 1b), all cells exhibited overall rounding but not in a monotonic fashion (note that spontaneous circularity features are central in multicellular behavior theory[17,25,34], but are experimentally revealed here for the first time; Extended Data Fig.2). In contrast, both expansion and contraction temporal responses during deadhesion in all examined tissues were immediate and monotonic (Extended Data Fig.3).

When we performed the deadhesion at different tissue ages, we found that the more mature and jammed the layer was (Fig. 2a), the larger was the fraction of expanding cells (Fig. 2b, Extended Data Fig.4). However, to our surprise, although tissue density plateaued during maturation, the fraction of expanding cells continued to rise (Extended Data Fig.5). Most intriguing was the fact that expanding cells tended to appear not as spatially random individuals but as clustered chains and islands that emerged and grew in cell numbers with maturation (Fig. 2a, b, Extended Data Fig.4).

To explore the extent to which clustering and expansion-contraction behavior is reflected in nucleus and cytoskeleton, we used confocal microscopy to identify apical actin rings, and superimposed cells' contours on the basal plane (Fig. 2c-d). The basal plane revealed two distinct actin phenotypes: phenotype A, with actin concentrated predominantly at the cell cortex, and phenotype B, with actin distributed throughout the cell interior. Notably, as confocal images were acquired post-fixation, we could not directly determine the mechanical phenotype (expanding vs. contracting) of individual cells in this analysis. However, what did aid was the examination of the compactness of the nuclei and the actin cytoskeleton in cells. The data show that expanding cells, relative to contracting cells, and phenotype B, relative to phenotype A, exhibit both a higher nucleus-to-cell area ratio and a smaller cell area (Fig. 2e, Movie S2, Extended Data Fig.6). Additionally, cells with similar actin phenotypes tended to spatially cluster and resemble the islands pattern observed in expanding and contracting cells (Fig. 2d, Extended Data Fig.7). Hence, the data suggest the following interpretation: expanding cells contain interior-based actin linked to a compressed structural state (Fig. 2f; phenotype B), whereas contracting cells contain cortex-based actin linked to a tensed structural state (Fig. 2f; phenotype A).

**Growth of compressed cell islands is reversible, detainable, and inhibits proliferation**

The expanding cell islands are seen to be embedded within an ocean of contracting cells. Such collective behavior in living epithelial cells is reminiscent of collective behavior of inert granular particles that are densely packed on a flat 2D surface. Among such particles, physical forces are transmitted through well-defined compressed regions that take the form of islands and chains[35,36]. The chains and islands vanish when external forces originating from the boundary of the system cease.

With that idea in mind, we generated a free boundary on the epithelium by creating a wound (Fig. 3a). From the edge of the wound, cells migrate forward to fill the void, and a wave of unjamming propagated retrogradely into the cell layer[10,37]. After cells migrated for a defined period, we performed the deadhesion assay and mapped the contracting and expanding phenotypes (Fig. 3b), together with the dynamics and morphology of each cell throughout its migration period (Fig. 3c). We then observed how a spatial gradient of migration velocity[10,37], directed toward the wound edge, coincided with a spatial gradient in the fraction of expanding cells, directed outward from the wound edge (Fig. 3c). The same behavior was observed for cell shape and area (Extended Data Fig.8). Together, these results demonstrate that unjamming and associated relaxation caused cells with an expanding phenotype (pre-relaxation) to transition to a contracting phenotype (post-relaxation). This means that the growth process of expanding cell islands stopped and reversed.

The relaxation patterns observed above (Fig. 3a-c; Extended Data Fig.8) were strikingly similar to those observed in force transmissions in inert granular systems. This is surprising for many reasons, not the least of which is that epithelial cells are soft and dynamically active, whereas grains are inert and virtually rigid. We wondered, therefore, about the extent to which these two collective systems still share further mechanical similarities. For example, as in the compressed granular regions, could expanding cells be in a state of compression? To approach this question, we considered two separate implications of compression and tension in epithelial systems.

First, if cells that were identified as expanding (post-deadhesion) were indeed more compressed, they might also interact (pre-deadhesion) with their immediate neighbors in a manner that reflects cell-cell repulsion. To explore this possibility, we defined a metric that quantifies local deformation (Fig. 3d, inset), thus reflecting the dynamic repulsion that each cell experiences with respect to its close neighbors during migration. Close to the wound's edge, both contracting and expanding cells immediately increased their cell-cell repulsion, and to a similar extent (Fig. 3d). However, far from the wound's edge, where the retrograde wave of unjamming had not yet penetrated, and overall collective motion did not yet exist, expanding cells displayed substantially higher repulsive behavior (Fig. 3d). We also observed how this behavior changes with the progression of time (Extended Data Fig.9).

Second, compression and tension in biological matter are translated to biochemical events and vice versa. We thus decided to examine the extent to which these observations are consistent with the known mechanobiological cascade of tensile forces that advance cell cycle progression[3-5,9,38]. This was done by mapping the cell cycle stage[8,9,39], in two opposing regimes: maturation, which reduces overall intercellular tension and naturally minimizes proliferation; and wound healing, which elevates tension and overall promotes proliferation. In these opposing cases, expanding cells were consistently less probable to advance in the cell cycle than contracting cells (Fig. 3e, f, g). That is, the expanding phenotype inhibits proliferation.

We then wondered about the opposite effect, namely, to what extent does cell-cycle re-entry affect the expanding cells network evolution? To answer this question this, we treated a mature and jammed tissue with Thymidine to arrest cell cycle progression and subsequent divisions[9]. Non-treated tissues showed a significantly bigger fraction of expanding cells than their Thymidine-treated counterpart, and more so, the treated tissues had maintained the same fraction as before the treatment (Fig. 3h). Hence, by inhibiting cell division, the growth process of expanding cell islands, and the evolution of the entire network that they form together (Fig. 2b), was detained.

Together, data presented thus far, namely, elevated repulsion in expanding cells and cell cycle activity in contracting cells, imply that contracting cells are cycling and tensed, and expanding cells are non-cycling and compressed.

**The compression network self-organizes in a biased percolation fashion**

The evolution of the compression network described above depended heavily on cell division (Fig. 3h). During advanced maturation stages, however, the fraction of compressed cells continued to grow, while the global tissue density remained constant (Extended Data Fig.5). Because no net density change occurred, the effects of each division must be local. This leads logically to the conclusion that cell divisions locally promote a compressed phenotype in a daughter cell or its neighbors. But in what fashion does this behavior contribute to the growth of the observed islands?

Given that cell-cycle re-entry predominantly occurs within a tensed region, a division event will promote the formation of a new compressed cell. This cell will either form a new compressed island or be annexed to an existing compressed island. Such outcomes might be thought to occur with no predisposition since cell-cycle re-entry is uniformly distributed in tensed regions (Fig. 3f). However, the early appearance of compressed cell islands (Fig. 2b, left) suggests a bias toward the growth of existing islands. This 'rich-get-richer' behavior resembles preferential attachment seen in many self-organizing systems—such as force chains in granular materials, forest fires, neural networks, and social systems—where new elements preferentially join larger clusters, driving scale-free organization[40-52].

Might such a biased self-organization mechanism contribute to the compression network evolution? To explore this possibility, we mapped adjacent compressed cells into well-defined islands that showed a wide variety of island sizes, $S$ (Fig. 4a). We then compared these data to a computational model that incorporates preferential attachment in the islands' growth rules. We used a minimal simulation of the epithelial layer represented as a hexagonal grid with an initial state that is fully occupied with "tensed" cells. At each step of the simulation a transition is made from a tensed to a compressed phenotype for one random cell. That cell is selected based on a probabilistic decision tree: either chosen randomly from the entire grid or, with a biased probability $p_b$, chosen from the immediate neighbors of an existing island (Extended Data Fig.10). When the latter occurs, the simulation favors attachment to larger islands, thereby implementing preferential attachment (Movie S3). This simulation captures the shapes and structure of the compressed islands qualitatively (Fig. 4b). To explore it quantitatively, we next examined the extent to which the epithelial compression network is characterized by a statistical distribution of island sizes that take a scale-free form[40]. A rigorous statistical analysis of island sizes revealed a characteristic fractal dimension (Supplementary figure *1*) and a clear scale-free power-law distribution (Fig. 4c; Supplementary figure 2). Importantly, that distribution was insensitive to fluctuations in expanding-contracting classification (Supplementary figure 3). Moreover, for all experimental conditions and all manipulations performed, the relationship between the proportion $P$ of compressed cells, and the power-law exponent $\alpha$, gave rise to a clear trend (Fig. 4d). That trend halted at $\alpha \approx 3/2$ and is clearly accounted for by the computational model.

A close examination shows that for $P \leq 20\%$ the experimental data points are generally below the $p_b = 0$ curve of the model. Specifically, these experimental data can be accounted for by the curves ranging between $p_b \approx 0.05 - 0.2$ (Fig. 4d). This suggests that $p_b$ might serve as a measure of the statistical preferential bias in the early stage of the network evolution. With the gradual increase beyond $P \approx 20\%$, the comparison to the model shows a diminished effect of $p_b$ (Fig. 4d).

It is important to emphasize that the model shows an increase in $\alpha$ at $P \approx 45\%$, and a non-monotonic behavior in $\alpha$ for $P > 45\%$. However, our experimental efforts have never yielded any significant crossing above $P = 45\%$. This discrepancy prompted us to consider whether the experimentally observed upper limit at $P = 45\%$, and stagnation at $\alpha \approx 3/2$ (Fig. 4d), might reflect an underlying self-organization process that is biologically bounded.

**In homeostasis the compression network is poised near criticality**

An increasing exponent $\alpha$ represents an increase in the overall proportion of small islands, and a decrease in the proportion of large islands. Accordingly, we suspected that the increase in $\alpha$ at $P \approx 45\%$ in the model (Fig. 4d) is evidence of a sudden behavioral shift, in which big islands start to merge with one another. To explore this idea, and its implication for our experimental data interpretation, we quantified the average island size, $\bar{S}$. The experimentally observed $\bar{S}$ showed an agreement with the model for up to $P \approx 45\%$, with an intensifying increase (Fig. 4e). Interestingly, that increase was independent of global cellular density, which remains constant for $P > 20\%$ (Fig. 4e, inset). However, compared to the experimental data, the model showed an even more dramatic increase in $\bar{S}$. Such a pronounced rise bears similarity to diverging correlation lengths in a second-order phase transition that might occur near our suspected critical point of $P = 45\%$ (Fig. 4e). To establish criticality at this point, we show its independence of lattice size $L$ – a well-established approach in studying continuous phase transitions near a critical point [17,53-55]. Specifically, we have identified the scaling ansatz (Eq.1)

$$\text{Eq.1:} \quad \bar{S}L^{-\gamma/\upsilon} = (P - P_c)^{\gamma} f\left(\frac{L^{-\gamma/\upsilon}}{(P-P_c)^{\upsilon}}\right)$$

on which all the model and experimental data perfectly collapse - a condition representing a critical behavior near $P_c \approx 45\%$ (Fig. 4f). For $P \gg P_c$, the results are consistent with the classical percolation threshold of 69% occupation, reported for an infinite hexagonal lattice [56,57]. Most importantly, and in a manner that is independent on $L$, we observed the following. For $P \ll P_c$, the lower branch of $f$ takes a linear form with $\bar{S} \propto (P - P_c)^{\gamma-\upsilon}$. And, as $P$ approaches $P_c$, the two branches of $f$ coincide and $\bar{S} \propto L^{\gamma/\upsilon(1-\gamma/\upsilon)}$.

Taken together, these results imply that the network of compressed cells tends to self-organize towards a critical point in which large islands start to merge (Fig. 4f, top inset). However, the network does not evolve beyond that point, and shows a halt both in average island size (Fig. 4e) and in $\alpha$ values of $\approx 3/2$ (Fig. 4d; Fig. 4f, bottom inset). Therefore, the compression network can be said to self-organize towards a quasi-critical state.

## Discussion

The epithelial compression network reported here is an emergent biological structure comprising islands of mechanically compressed and non-cycling cells, embedded within an ocean of mechanically tensed and cycling cells. Importantly, these cell-cycle patterns continue to evolve in space and time even as cellular density and subsequent cell-cell contacts approach a steady state (Extended Data Fig.5, Fig. 3e,f). This dynamic behavior demonstrates how the local mechanism of contact inhibition of proliferation (CIP) is subsumed within the non-local mechanism of network-based inhibition of proliferation (NIP).

This epithelial compression network evolves towards a quasi-critical state. Such behavior is described as self-organized criticality (SOC) – when a system self-tunes toward a critical point and produces long-range correlations, scale invariances, and power-law distributions[42,58]. All these behaviors are expressed by the epithelium compression network, but also expressed by inert systems including sandpiles[59,60], earthquakes[61], and forest fires[62], and by other biological systems such as neurons[45-51], metabolites[43], and actin-myosin suspensions[63]. In these biological systems SOC has been linked to optimization of functionality, and although the implication of SOC in an epithelial compression network remains unclear, some clues are suggested here. Upon wounding, for example, mechanical communication radiates from the wound edge as spatial gradients of motion, morphology, and fraction of compressed-cells (Fig. 3a–c, Extended Data Fig.8). Notably, out of the three, the gradient in the fraction of compressed cells propagates farther and faster. This finding suggests the possibility of mechanical events – transmitted through chains and islands of compressed cells – serving as an efficient means of signaling and coordinating wound repair. If so, then the specific critical structure ($\alpha = 3/2$) might be a factor providing optimal long-range transmission of those mechanical signals while keeping the surrounding tissue tensed and primed for cell-cycle re-entry after injury.

These findings have potential implications for tissue development and cancer growth. In development, spatial patterns of proliferation and compression may guide morphogenesis, while in oncogenesis the compression network may confine or permit physical pathways for tumor growth. Thus, important and unanswered questions include to what extent, and in what ways, the compression network is dynamically modified in disease. In this context, the physical picture presented here captures not only the ultimate critical state but also the dynamic pathway leading to it (Fig. 4).

Accordingly, the discovery of the epithelial compression network and its evolving critical structure adds a new dimension to our understanding of tissue homeostasis. It implies that the homeostatic state is not merely a static endpoint of proliferation arrest and motility loss, but rather an ongoing dynamic and spatially heterogeneous state balanced at the edge of criticality.

**Acknowledgements:**

We thank and acknowledge helpful discussions with Dapeng Bi, Eyal Karzbrun, Jeffery Fredberg, Yaniv Edery, and David Weitz. L.D. acknowledges financial support from the Israel Scholarship Education Foundation (ISEF). L.A acknowledge financial support by the Israeli Science Foundation (ISF) on the individual research grant number 2107/21, and the New-Faculty Equipment grant number 2108/21; Israeli Ministry of Innovation, Science and Technology 0357-24; BGU's joint collaboration grant, Engineering Sciences and the Blaustein Institutes for Desert Research, 2025; Pearlston Center at BGU; Israeli Young Academy

**Author contributions:**

L.D., Y.F., and L.A. conceptualized the research. L.D., Y.L., and L.A. designed the experiments. L.D., A.S., and Y.L. performed the experiments. L.D and D.S developed the post-deadhesion tracking analysis algorithm. L.D implemented the simulations on the hexagonal lattice. L.D designed and performed all presented analyses. L.D., Y.L, Y.F. and L.A. contributed to data interpretation. L.D and L.A. wrote the manuscript. L.A oversaw the project.

**Competing interests:**

Authors declare that they have no competing interests.

**Data and materials availability:** All data are available in the main text or the supplementary materials. Specific codes used will be shared upon reasonable request.

## Figures

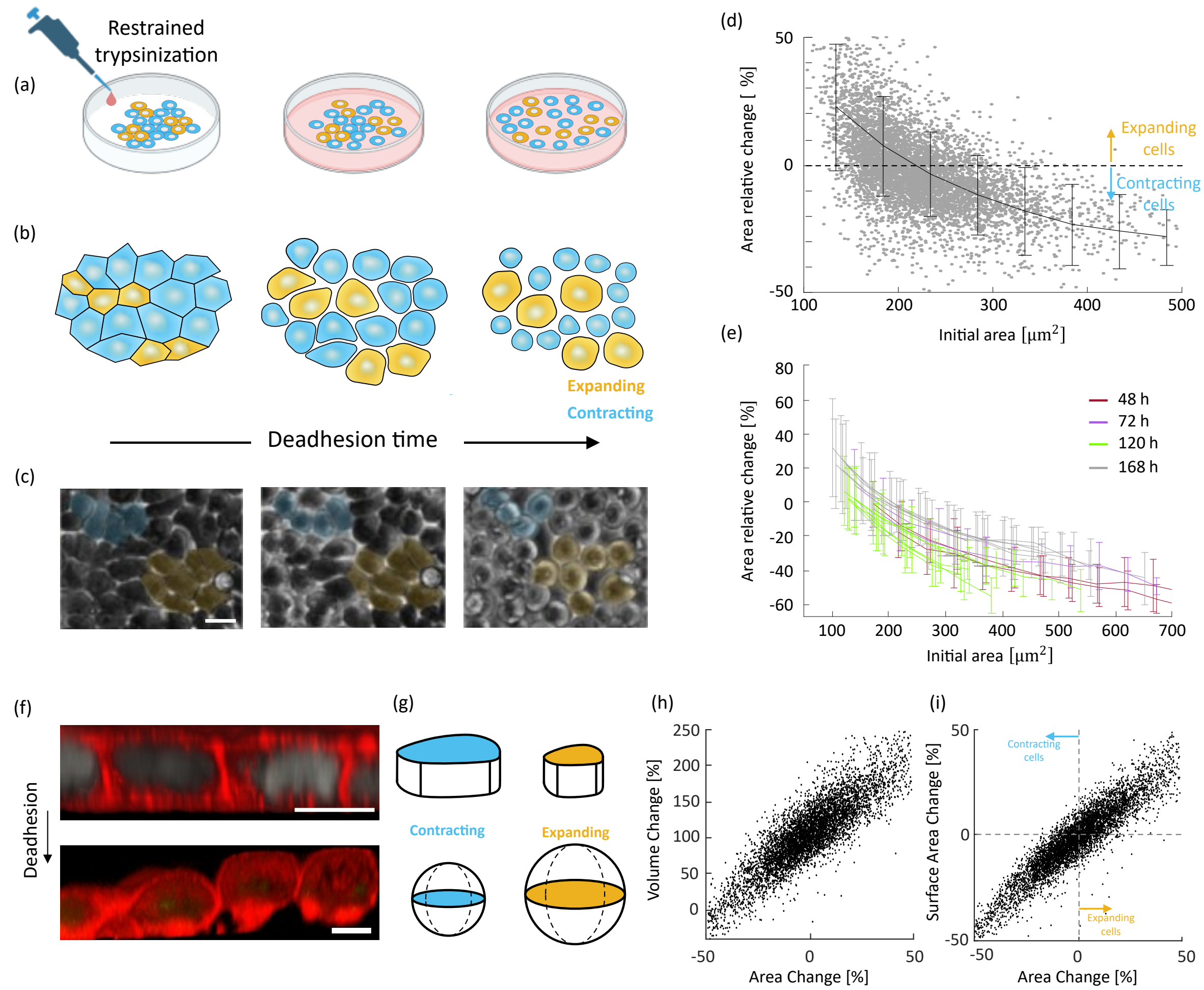

**Fig. 1 - Enzymatic deadhesion reveals two mechanically distinct cell populations - contracting vs expanding.** (a) A restrained trypsinization protocol was applied to mature MDCK monolayers to eliminate cell-to-substrate and cell-to-cell adhesion. (b) Cell deadhesion reveals two distinct mechanical responses that were observed through **2D inspection**: cell expansion and cell contraction. (c) Time-sequential snapshots of the deadhesion process showing contracting cells, marked in blue, and expanding cells, marked in yellow (scalebar 15 µm). (d) The area-relative-changes of both contracting and expanding cells are correlated with their initial area, with a Pearson coefficient of $-0.55 \pm 0.02$ for 168 [h] matured tissues, and with 5 different field of views (FOVs). The scatter plot shown contains 6097 cells from one representative FOV, and the error bars represent SD. (e) The area-relative-change to initial-area correlations systematically persisted throughout different stages of maturation, with Pearson coefficient of $-0.64 \pm 0.07$, $-0.55 \pm 0.06$, $-0.61 \pm 0.08$, $-0.55 \pm 0.02$ for 48, 72, 120, and 168 [h], taken from 4, 11, 9, and 5 different FOV, respectively. A FOV contains 3345 cells on average. The data was collected from n=10 independent experiments. (f) For **3D analysis**, we used a 168 [hr] matured tissue grown on glass and fixed. Shown are confocal imaging sections for a non-trypsinized layer with a column-like cell geometry (top), and a trypsinized layer with a sphere-like cell geometry (bottom). Both scale bars represent 7 µm. (g) Illustration of the 3D shape transition for expanding and contracting cells. (h) Measurements of 3D volume changes indicate that both expanding and contracting cells experience an increase in volume. (i) Measurements of 3D surface area change reveal an opposite response between the phenotypes.

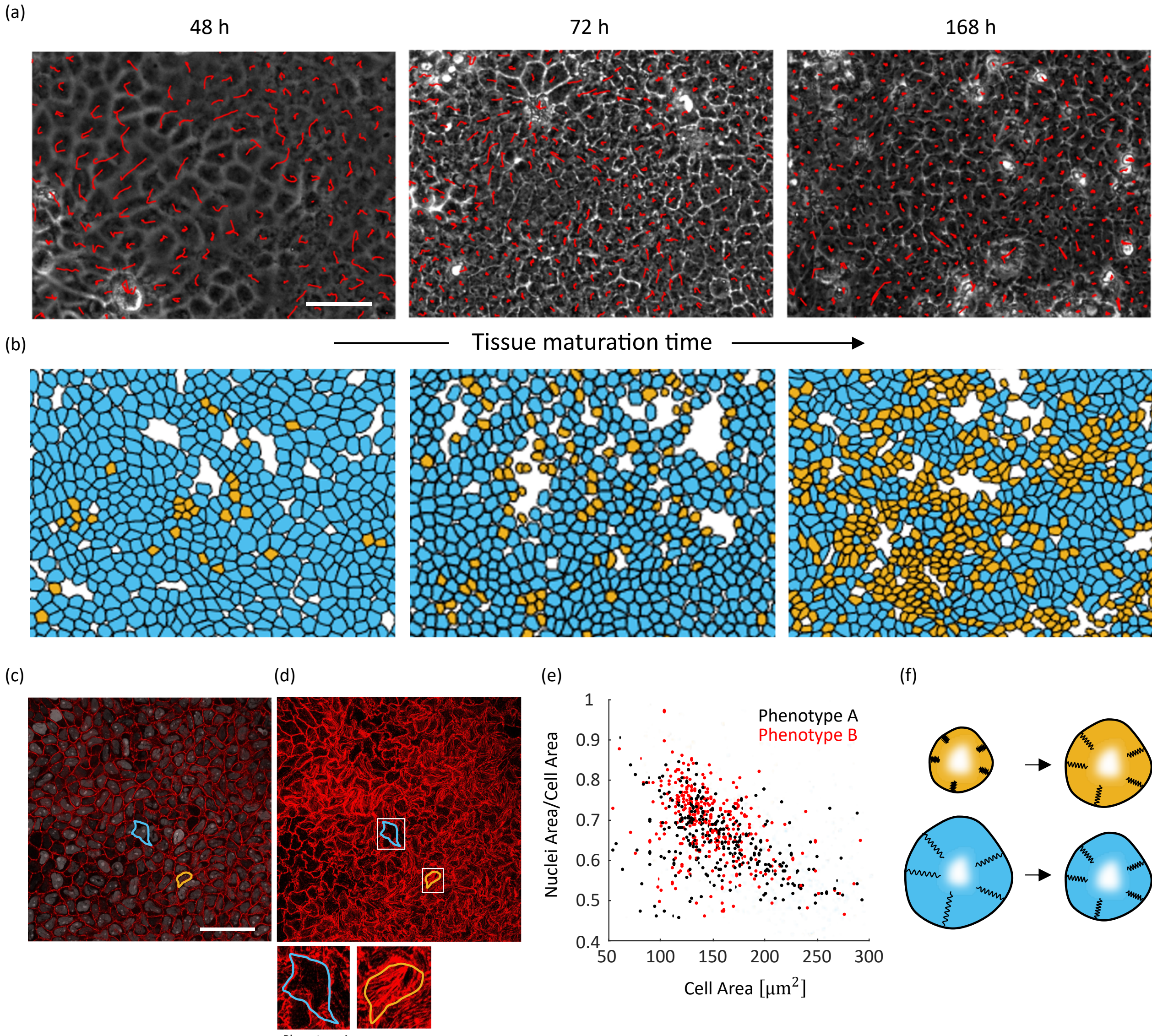

**Fig. 2 - Islands of expanding cells emerge and grow in cell numbers with gradual jamming, and display distinguishing features by both nucleus compactness and spatial arrangement of the actin cytoskeleton.** (a) Gradual jamming and reduction in motility throughout maturation are reflected in the gradual shortening of trajectory curves. Each curve is based on 200 minutes of tracking (scalebar 50 μm). (b) Color-coded maps of expanding and contracting cell phenotypes in pre-trypsinized tissues at various maturation stages. Expanding cells are marked in yellow, and contracting cells in blue. White spaces represent cells not recognized by the segmentation algorithm (scalebar 100 μm). Color coding reveals growing islands of expanding cells, embedded in an ocean of contracting cells. (c) Confocal image showing both nuclei, and actin rings that represent cell contours. Highlighted are two examples of identified contours (scalebar 50 μm). (d) The two exemplifying contours are projected onto the basal plane. The basal plane shows two distinct actin arrangements: phenotype A, in which actin is primarily localized at the cell perimeter; and phenotype B, in which actin is distributed throughout the cell interior. Actin arrangement analysis was done for 168 [h] tissue with F-actin and nuclei markers. (e) For both actin phenotypes, as well as for expanding and contracting cells, nucleus compactness (area ratio between nucleus and cell) is presented as a function of the cell's initial area. Statistical analysis shows similarity between contracting cells and actin phenotype A, as well as between expanding cells and actin phenotype B (Extended Data Fig.6). This similarity is reflected by higher nucleus compactness and smaller cell area in both expanding cells and actin phenotype B cells. (f) These observations, together with those in Fig. 1i, suggest that the membrane-to-nucleus elastic network is pre-compressed in expanding cells and pre-tensioned in contracting cells.

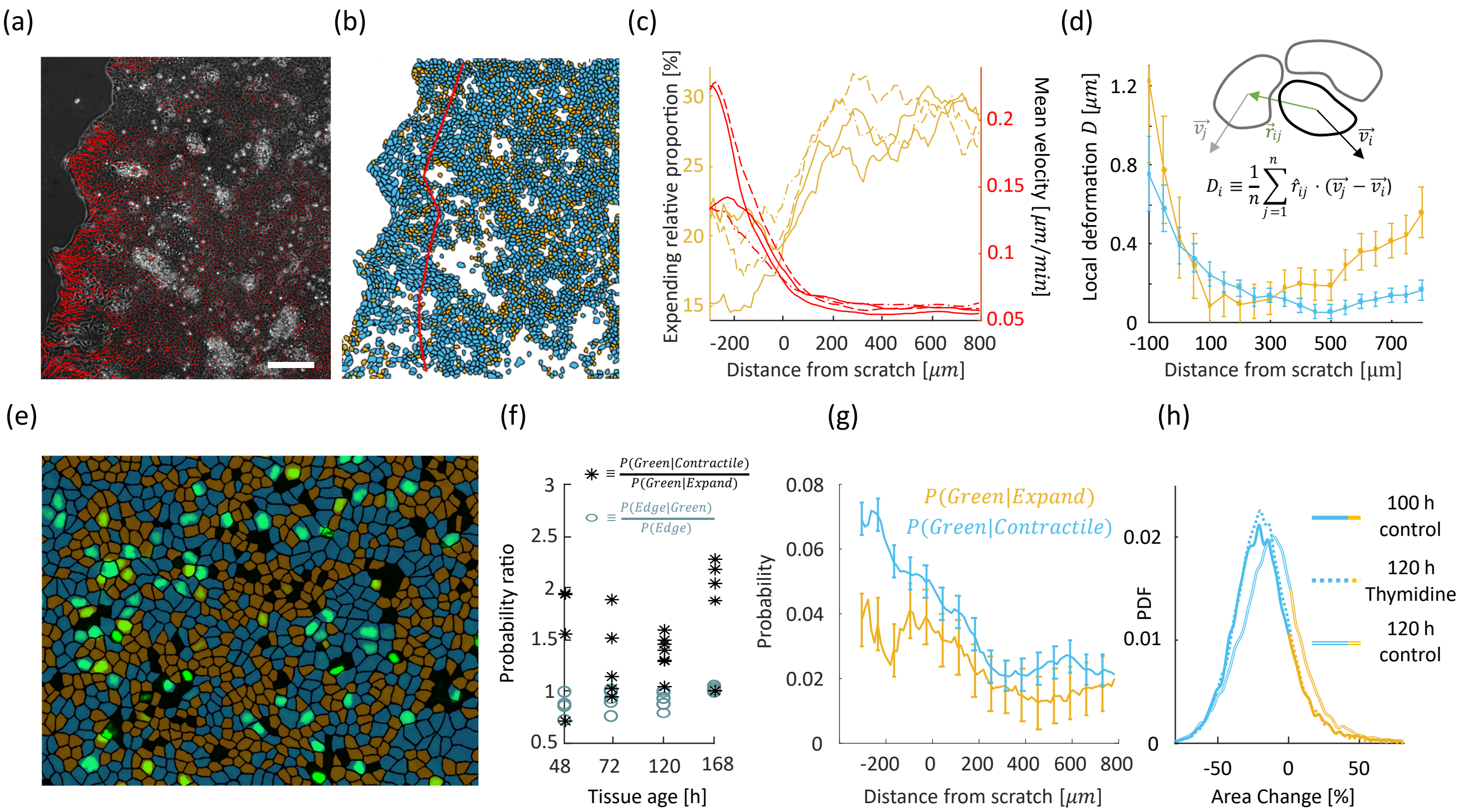

**Fig. 3 - Growth in the number of expanding cells is reversible through unjamming transition, detainable by cell-cycle arrest, and leads to specific spatial inhibition of proliferation.** (a) A scratch was inflicted on a 168-hour mature, confluent, and jammed tissue. The unjamming transition is reflected by a gradual elongation of the trajectory curves as the cells position is closer to the migrating wound edge (trajectories tracked over 200 minutes, scale bar 150 $\mu m$). (b) After 11 [h] of migration period, we performed the deadhesion assay and mapped the contracting (blue) and expanding (yellow) phenotypes. Red curve represents the initial scratch position. (c) Relative proportion of expanding cells (yellow curves) and migration velocity (red curves) were calculated with a moving average at various distances from the leading edge. Averaging was done within a 276 μm-wide strip at each distance, with different curve types corresponding to different FOVs. (d) Local tissue deformation, which reflects cell-cell repulsion, was calculated immediately after the unjamming scratch. Average deformation around cell $i$ was calculated with respect to its $n$ nearest neighbors (inset). Deformations are shown for 10 minutes after wounding, with later times shown in Extended Data Fig.9. (e) Cells in S/G2/M stage in the cell cycle were observed using the FUCCI's green fluorescent reporter (EGFP) that was overlaid on the expanding-contracting map. (f) Statistical analysis of cell cycle progression during maturation. Asterisks - representing the ratio of the probability that a cell possesses a green signal given that it is a contractile cell, to the probability that it possesses a green signal given that it is an expanding cell; Circles - representing the ratio of the probability that a cell is located near an island edge given that it possesses a green signal, to the probability that it is located near an island. (g) We repeated the cell-cycle progression analysis during wound healing, reporting probabilities as a function of distance from the wound edge. Data are shown as mean and SD, averaged across 4 FOVs. (h) Histograms displaying the PDFs of post-deadhesion relative area change for: 1) Tissue matured for 100 hours (data was taken from 11 FOVs); 2) Tissue matured for 120 hours, pre-treated with Thymidine for 3 hours at 100 hours (data was taken from 9 FOVs); 3) Tissue matured for 120 hours (data was taken from 7 FOVs).

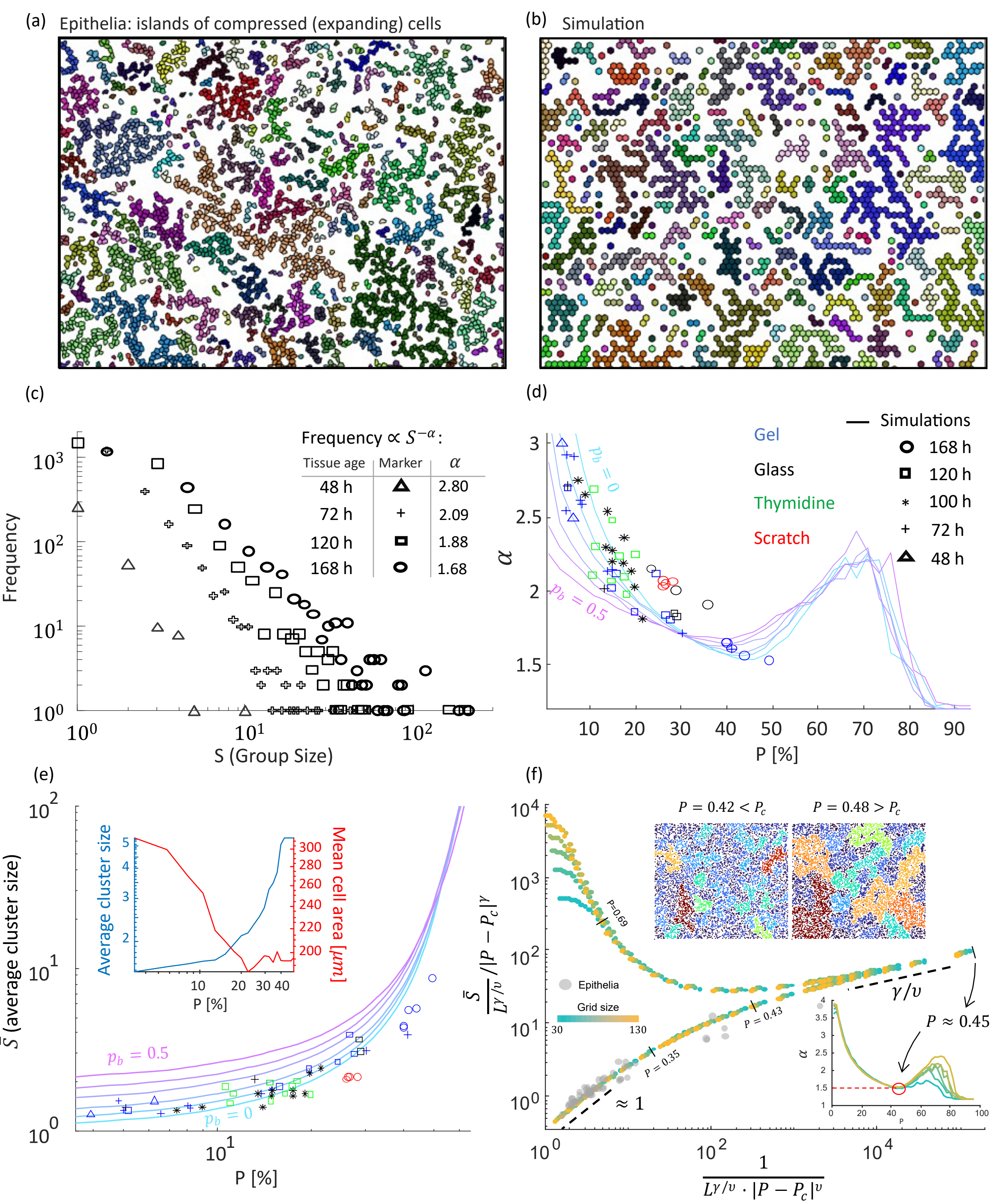

**Fig. 4 - Islands of compressed (expanding) cells evolve through a biased percolation-like process that approaches a quasi-critical point, and avoids mergers of large islands.** (a) Compressed cells in a 168 [h] matured tissue, colored to distinguish different islands. Islands were classified based on a spatial proximity criterion. (b) Simulated islands were generated by a computation model for site percolation on a hexagonal grid, which incorporates preferential attachment bias (Extended Data Fig.10; Movie S3). (c) Island size distributions in epithelia follow a power-law, with frequency ∝ $S^{-\alpha}$ , as reflected by the linear trend in the log-log plots. The slope represents the power-law exponent α, fitted using an iterative scheme (Supplementary figure 2). Data were collected from 4 FOVs for 48 h, 11 FOVs for 72 h, 9 FOVs for 120 h, 5 FOVs for 168 h. (d) The power-law exponent α is plotted against the proportion of expanding cells P, for both experimental data and biased percolation simulations. While the experimental

results evolved only up to $P \approx 45\%$, simulation data were analyzed over the full range of occupation proportions P. Scatter data colors represent experimental conditions. Solid lines colors represent simulations that were performed with bias probability values $0 \leq p_b \leq 0.5$, and averaged over 29 simulations (Extended Data Fig.10). (e) Average island size (in number of cells) is plotted against P for both simulated and experimental datasets. The inset shows the experimental island size (blue) alongside mean cell area (red) plotted against P, and highlights a continuous increase in island size, despite tissue density reaching a constant and steady value. (f) Scaling analysis (with $p_b = 0$) reveals a critical point at $P_c = 45\%$, where large islands begin to merge (inset, top) and the power-law exponent approaches 3/2 (inset, bottom). Ansatz's parameter values are $\gamma = 0.45$ and $\upsilon = 1.83$. The same ansatz was applied to the experimental data (grey circles), which exhibited the same trend as the simulations, both qualitatively and quantitatively. For consistency with the simulations, the characteristic length $L$ for the experimental data was defined as the square root of the number of all identified cells in the FOV.

## Materials and Methods

### Polyacrylamide gel (PAG) preparation

Glass-bottom dishes were activated by coating with a 12:1:1 solution of 99% ethanol/acetic acid/bind silane (M6514, Merk) for 30 minutes, washed twice with 99% ethanol, and dried. 0.5 ml PAG solution (30 kPa) was prepared from 310 $\mu L$ PBS (D405, Lifegene), 150 $\mu L$ acrylamide (1610140, BioRad), 37.5 $\mu L$ bis (1610142, BioRad), 2.5 $\mu L$ APS (1610700, BioRad), and 0.25 $\mu L$ TEMED (T9281, Merck). PAG solution was deposited as 20 μl drops in activated dishes, overlaid with 16 mm coverslips, and polymerized for 1 hour. After adding PBS, coverslips were removed, and gels were UV-cured and sterilized. For cell adherence, gels were activated with 50 $\mu L$ SANPAH (PHC-c1111, Proteochem), washed until clear, and coated overnight with 0.1 mg/ml collagen (5005, BioTag) in HEPES (CA-25-060-CI, Getter-Biomed) before PBS wash.

### Cell culture and microscopy

Madin-Darby Canine Kidney (MDCK) II cells stably expressing the FUCCI (Fluorescent Ubiquitination-based Cell Cycle Indicator) transgene were kindly provided by Dr. Lars Hofnagel[8]. Cells were maintained in Modified Eagle's Medium (MEM; Merck) supplemented with 10% fetal bovine serum (FBS; Merck), 1% penicillin–streptomycin (Pen-Strep; Merck), and 1% L-glutamine (Sartorius). Cultures were incubated at 37 °C in 85% humidity and 5% $CO_2$. Cells were seeded onto either PAG, or directly on glass at a density ranging from 700-1200 cells/mm$^2$. Cells were cultured for 48, 72, 100, 120, or 168 hours.

### Trypsinization protocol – restrained deadhesion

In most cases, when trypsinization is performed for routine culturing purposes, the majority of the tissue is elevated from the substrate in large, connected groups (Supplementary figure 4). To avoid this, we have customized the conventional trypsinization assay to allow some of the medium to stay trapped in extracellular spaces. Thus, when trypsin was added its effect was slightly inhibited and the deadhesion process was restrained. We started by first washing each dish, by tilting and repeated pipetting, with warm (37°C) culture medium. The dish was then positioned on a leveled surface, and the medium was aspirated from the edge of the dish, far from the tissue, and in a manner that did not remove all medium completely. We then added 1 ml of PBS at the edge of the dish and immediately aspirated it. The dish was then mounted onto the microscope stage (Axio-obserever 7, Zeiss) and imaging positions were set. We then added 500 μl of warm (37°C) 0.25%

Trypsin EDTA (D705, Lifegene)), from the edge of the dish, and immediately started the deadhesion imaging at 1 minute intervals, with 10x objective.

### Thymidine protocol

Cells were cultured as for restrained trypsinization experiment. At 100 hours, dishes were washed thoroughly with warm (37°C) PBS, then supplemented with 100mM Thymidine (T9250, Merck), in growth medium. Dishes were incubated with thymidine for 3h, washed, and then supplemented with growth medium and put back in the incubator until 120h passed from seeding. Treated dishes were then trypsinized at 120hrs.

### Time-lapse fluorescent and phase-contrast microscopy

Petri dishes were secured under a microscope with a stage top incubator (Ibidi), maintained at 37C, 85% humidity and 5% $CO_2$. Using 10x objective, 3 channel images were obtained (phase contrast, green and orange), every 10 minutes. Green channel was obtained with the 38 HE filter set (470/525, Zeiss). Orange channel was obtained with the 43 HE filter set (550/605, Zeiss).

### Confocal imaging

For confocal imaging, cells were grown on glass and fixed with 4% formaldehyde (TS-28908, Rhenium). After fixation, cells were permeabilized with 1% Triton x (HFH10, Rhenium), and blocked with 1% BSA (TS-37525, Rhenium). Subsequently, cells were stained with phalloidin Alexa fluor 647 (A22287, Invitrogen). Stained cells were imaged with Zeiss LSM880 airyscan.

### Green FUCCI detection and cell trajectories

Using a pretrained Cellpose model[39], we performed nuclei segmentation on the enhanced green fluorescent protein (EGFP) FUCCI channel. The FUCCI masks were analyzed together with the corresponding phase-contrast images, to identify cells that contained active FUCCI reporter. Additionally, leveraging the accuracy and accessibility of the FUCCI channel segmentation, we used TrackMate to generate cell trajectories for our time-lapse and scratch experiments.

### Data extraction

Despite following our restrained trypsinization protocol, in some cases, the deadhesion process was not suitable for analysis. Therefore, we focused on analyzing experiments where we achieved a gentle separation of individual cells (Supplementary figure 4). A gentle separation allowed us to accurately identify each cell boundary and track it over time.

We used Cellpose and TrackMate in FIJI to generate cell masks and extract the coordinates of the cell outlines[64-66]. A spline function was applied to these coordinates to avoid pixelated contours, enabling more accurate geometric measurements. Changes in cell geometry were analyzed by tracking the cell outlines over time across sequential frames with TrackMate (Supplementary figure 5).

To verify our results in general, and the cell expansion-contraction results in particular, we formulated a data examination protocol. In this protocol we created an algorithm which displays series of four sequential time frames and highlights a cell which TrackMate identified as the same cell. We then visually determined whether the cell had been tracked correctly or not. In Supplementary figure 6, we show examples of four tracking examinations with two examples of successful tracking and two examples of unsuccessful tracking. For quantitative assessment, we visually examined one field of view from which the algorithm generated ~6500 tracks. From these tracks, we first subsampled 179 tracks with a relative area change of -30% to 30%, as most the data is spread in this range. For this range, our examination showed virtually no tracking errors, with 1.68% of incorrect tracks. We then subsampled 1336 tracks in the extreme relative area change values that are lower than -30% or greater than 30%. For that range, we got 12.87% of incorrect tracks.

3D shape analysis

Based on our confocal images, it is reasonable to associate a column-like shape to cells in a mature confluent tissue, and sphere-like shape after complete separation of cells post deadhesion. In addition, confocal imaging allowed us to measure the tissue height, as we found a mean height of 7.2 μm (Extended Data Fig.1). Based on the above, we evaluated the cells' volume and surface area using both confocal and 2D phase-contrast microscopy. Hence, for cells in mature confluent tissue, we calculated cell volume using $V_{initial} = A_{initial} \cdot h_{measured}$ and cell surface area $S_{inital} = 2 \cdot A_{initial} + P_{initial} \cdot h_{measured}$. For post deadhesion cells, approximated as spheres, we calculate $V_{final} = \frac{4}{3}\pi r^3$ and $S_{final} = 4\pi r^2$, where $r$ is derived from the projected 2D circle in phase-contrast micrographs.

Groups detection

To investigate the size distribution of expanding cell islands, we utilized the binary expansion–contraction maps as a basis for identifying spatial groupings of expanding cells. A custom C++ program was developed to classify expanding cells into distinct groups based on spatial proximity.

Two expanding cells were assigned to the same group if the Euclidean distance between their centers was smaller than the sum of their equivalent radii (i.e., the radii of circles with areas equal to those of the individual cells), multiplied by a distance scaling factor. This factor was introduced to account for cases in which elongated cells are adjacent along their long axis. A distance factor of 1.4 was selected, as it yielded robust group identification based on visual validation of the output.

### Power law analysis

The group detection revealed that cell islands grew in size over maturation time. To better understand the clustering behavior of the expanding cells, we focused on analyzing the statistical distribution of group sizes. Observations indicated that group sizes increased in a manner resembling preferential attachment growth, a common mechanism in networks development. Such a mechanism is known to produce power-law distributions in the resulting population. Therefore, we attempted to fit our data into a power-law distribution and compared the goodness of fit against an alternative exponential distribution[67]. To fit a power-law distribution to our island size data, we initially considered maximum likelihood estimation (MLE) [67]. However, due to the limited size of our dataset, MLE produced biased results. Instead, we implemented an optimization-based numerical scheme, using the Kolmogorov-Smirnov (KS) statistic between the empirical and hypothesized model CDFs as the objective function. This approach allowed us to estimate the power-law exponent by minimizing the KS distance, providing a robust fit despite a small sample size. The scheme goes as follows: for a given dataset originated from an unknown distribution, we hypothesized that the data follows a power-law distribution $P(x) \propto x^{-\alpha}$ , where the exponent $\alpha$ lies within the range $1.2 < \alpha < 4$. To find $\alpha$ that best describe the data, we employ the two-sample Kolmogorov-Smirnov (KS) test. First, the empirical cumulative distribution function (CDF) of the dataset, $F_{real\ data}$ , is calculated. Next, synthetic data is generated based on the power-law distribution $P(x) = c_i x^{-\alpha_i}$ ($c_i$ being a normalizing factor depending on $\alpha_i$) for a specific value of $\alpha_i$ within the defined range, and the CDF of the synthetic data $F_{synthetic}$ is determined. The KS statistic, $D_{\alpha_i} = \sup\left(\left|F_{real\ data} - F_{synthetic}\right|\right)$, is then computed to quantify the maximum difference between the two CDFs. This process is repeated 15 times for $\alpha_i$ to calculate the average $\overline{D}_i = \frac{1}{15}\sum_{k=1}^{15}\left(D_{\alpha_i}\right)_k$. The procedure is repeated for multiple values of $\alpha_i$ within the range $1.2 < \alpha < 4$, and the fitted alpha will be the one which produces the lowest averaged $\overline{D}_i$.

To estimate errors, we applied the above numerical scheme to find the best power-law fit for synthesized data generated from a known distribution. These simulations were repeated for various known power-law exponents $\alpha$ and different sample sizes, as sample size is expected to influence the accuracy of determining the original distribution. Additionally, it is important to note that, in practice, the distribution of cell group sizes follows a discrete power-law that is limited in range. Consequently, when synthesizing data, we must define a cut-off for the maximum value that can be generated, which determines the value of $c$ in $P(x) = cx^{-\alpha}$. When analyzing our experimental data, we observed that the choice of cut-off was not particularly critical. However, for consistency, we set the cut-off to match the number of cells in the relevant field of view, as this represents the largest possible group size. Supplementary figure 7 presents an error analysis for various known $\alpha$ values (1.5, 2, 2.5, 3) across a wide range of sample sizes. The results show small and non-biased errors.

<u>Goodness of fit</u>

While our algorithm effectively determines the best-fit $\alpha$ for a power-law distribution hypothesis, it does not provide a direct measure of the statistical significance of the fit, specifically compared to alternative hypotheses. To determine whether the experimental data indeed follows a power-law distribution, we performed a statistical goodness-of-fit test. First, we estimated the power-law exponent $\alpha$ that best fits the empirical data. We then computed the KS statistic, $D_{data\ vs\ analytical}$, between the empirical cumulative distribution function (CDF) and the analytical power-law CDF corresponding to the fitted $\alpha$. To evaluate the significance of this fit, we generated 500 synthetic datasets from the fitted power-law distribution, and calculated their KS statistics against the same analytical CDF. The P-value was defined as the fraction of KS statistics exceeding $D_{data\ vs\ analytical}$ [67]. In most cases, the synthetic datasets yielded higher KS statistics than the empirical data, indicating that the power-law model provides a statistically robust fit. This conclusion was further supported by a comparison with exponential fits, which showed lower agreement (Supplementary figure 2).

153 **Extended data figure**

(a)

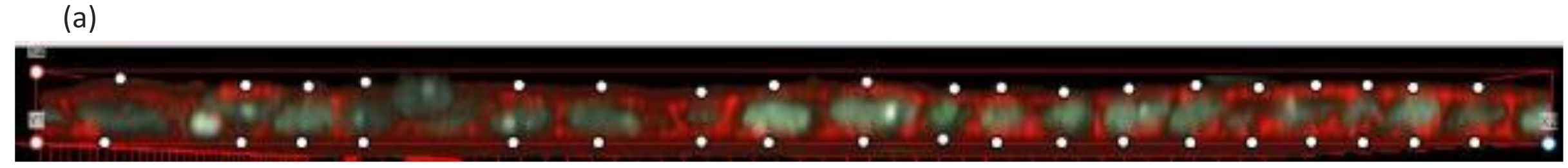

(b)

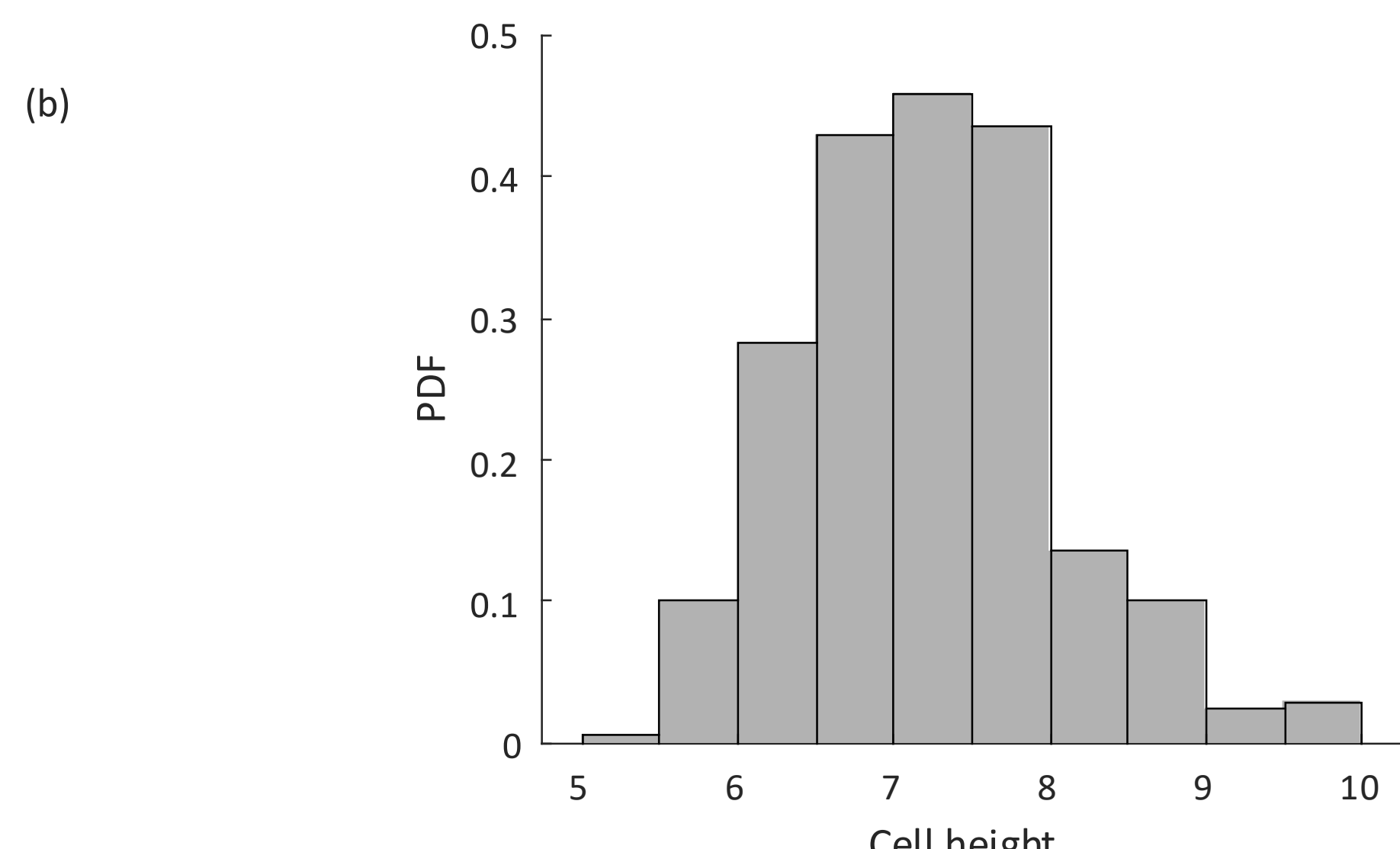

**Extended Data Fig.1 - Distribution of height measurements in 168 [h] matured tissue.** (a) Example of
a confocal cross-section in which we measured tissue heights. Vertical distance between couple of marked
points is considered as one data sample for tissue height. (b) The dataset includes 340 height measurements
collected from three different tissues, showing a narrow height distribution with an average of 7.2 $\mu m$.

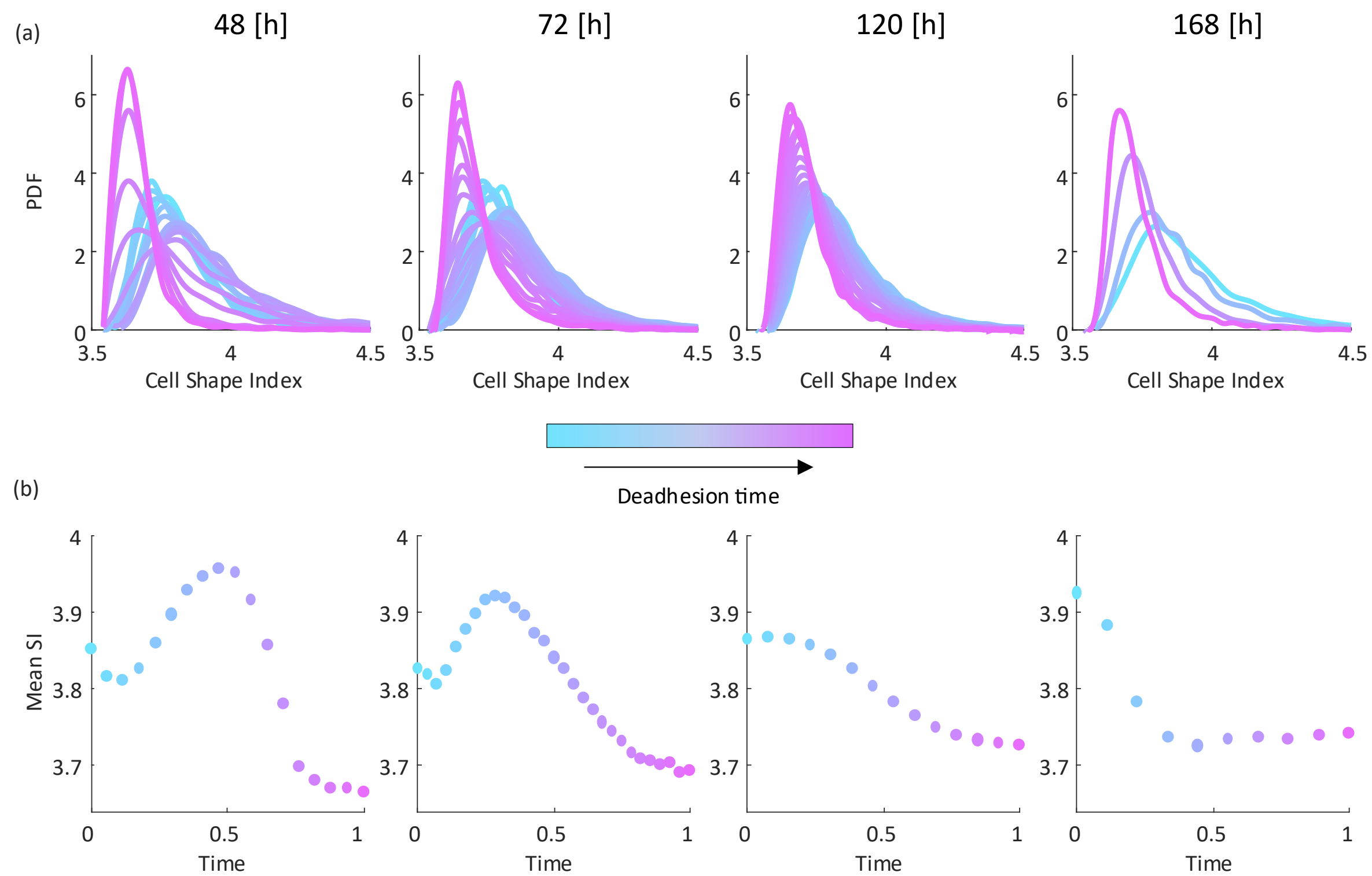

**Extended Data Fig.2 – Post-deadhesion cell rounding is observed in tissues at all stages of maturation,**
**however, in less mature and more motile tissues cells first go through an initial elongation stage.** The
concept of spontaneous shape plays a major role in interpreting collective behaviors in dense epithelial
tissues, specifically, the liquid-like to solid-like jamming transition that is predicted by the framework of
the vertex model [17,25,34]. The vertex model takes, as a central input parameter, the stress-free spontaneous
2D shape index (SI = $p/\sqrt{A}$ , $p$ – perimeter, $A$ – area) of cells. However, the nature of that spontaneous
geometrical shape is completely unknown experimentally. The deadhesion experiments in this study
revealed that all cells ended up rounding in the deadhesion process, but with distinct temporal features that
varied between tissue maturation stages (a) Histograms of the cell SI at different tissue maturation stages
(48–168 h). Each curve represents a single snapshot during the deadhesion process, which typically takes
15$\pm$7 minutes, with color coding reflecting normalized time progression from start (blue) to end (magenta).
Normalized time is defined as $T = \frac{t-t_0}{t_{final}-t_0}$ , where $t_0$ marks the onset of shape change, and $t_{final}$ marks
the point of stabilization. (b) Mean shape index (SI) plotted as a function of normalized time.

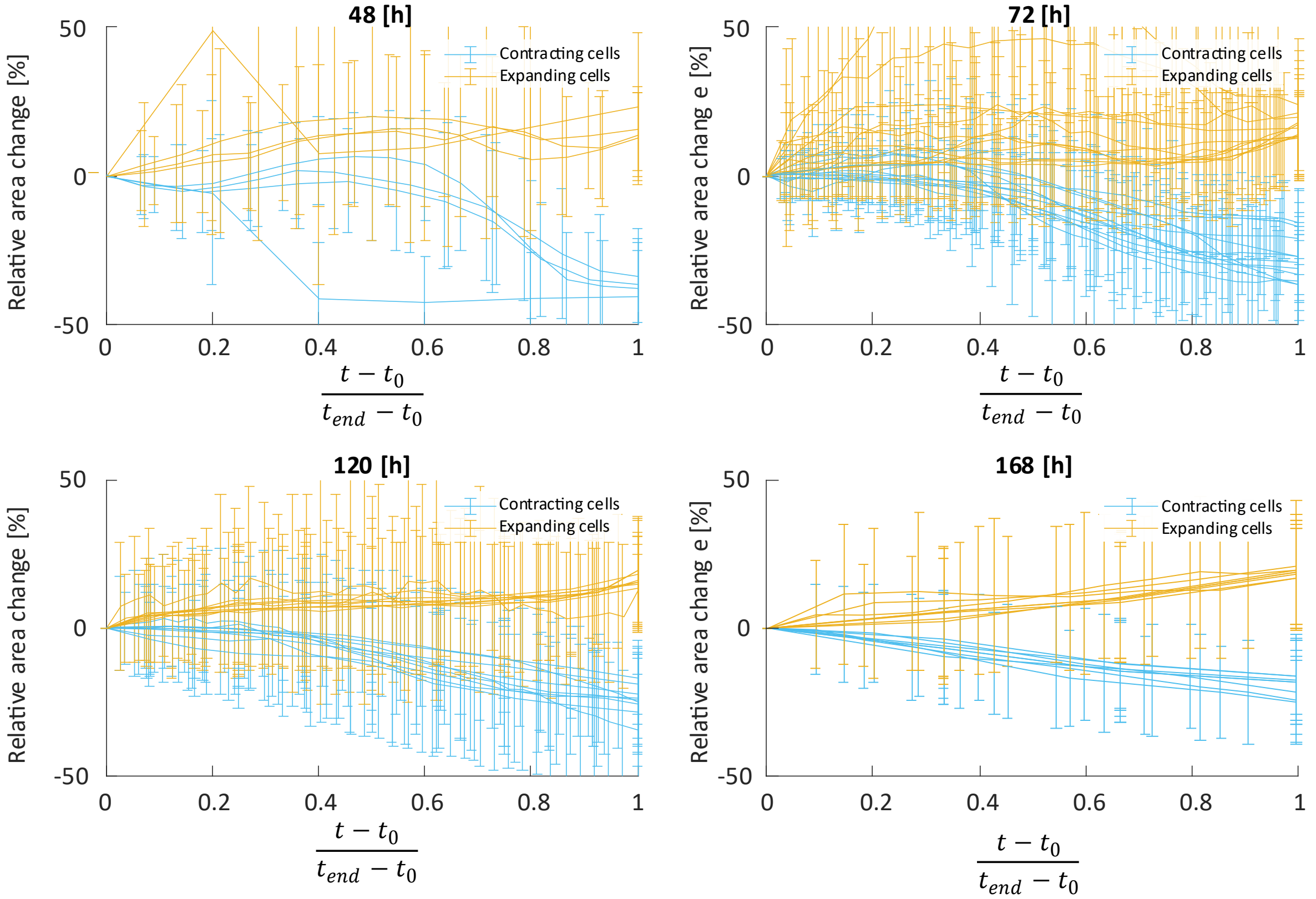

**Extended Data Fig.3 - Relative area changes of expanding and contracting cell populations were**
**overall consistent throughout the entire deadhesion process.** Time is normalized by the total duration of
the deadhesion process, which typically takes 15±7 minutes. Trends indicate that overall, cells classified
as expanding/contracting begin to increase/decrease in area immediately upon deadhesion initiation and
maintain that behaviour throughout the entire process. Trends consist across different tissue maturation
stages – 48, 72, 120, 168 [h], and include tissues grown on both glass and PAG. For all plots, each curve
represents one FOV, and the error bars represent the standard deviation of the data within that FOV.

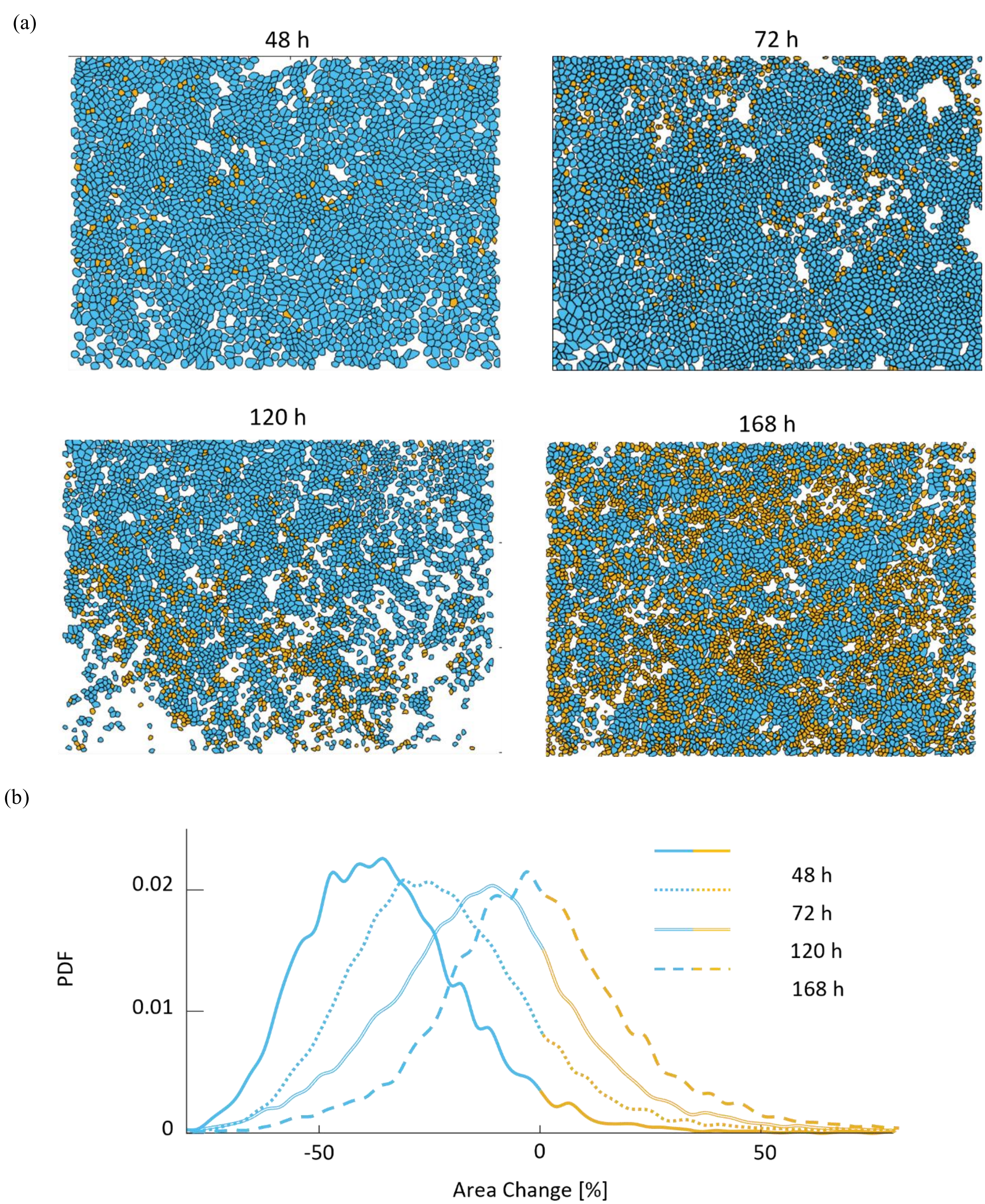

**Extended Data Fig.4 - Expanding cells grow in numbers and emerge in islands.** (a) Examples of
expanding and contracting maps across different tissue stages (48, 72, 120, and 168 h). Expanding cells are
shown in yellow, contracting cells in blue, and white spaces indicate cells not detected in the image analysis.
(b) PDFs of cells relative area change within different tissues aged 48, 72, 120, and 168 h. Curves containing
8566, 28033, 32687, and 27733 cells from 4, 11, 9, and 5 FOVs, respectively.

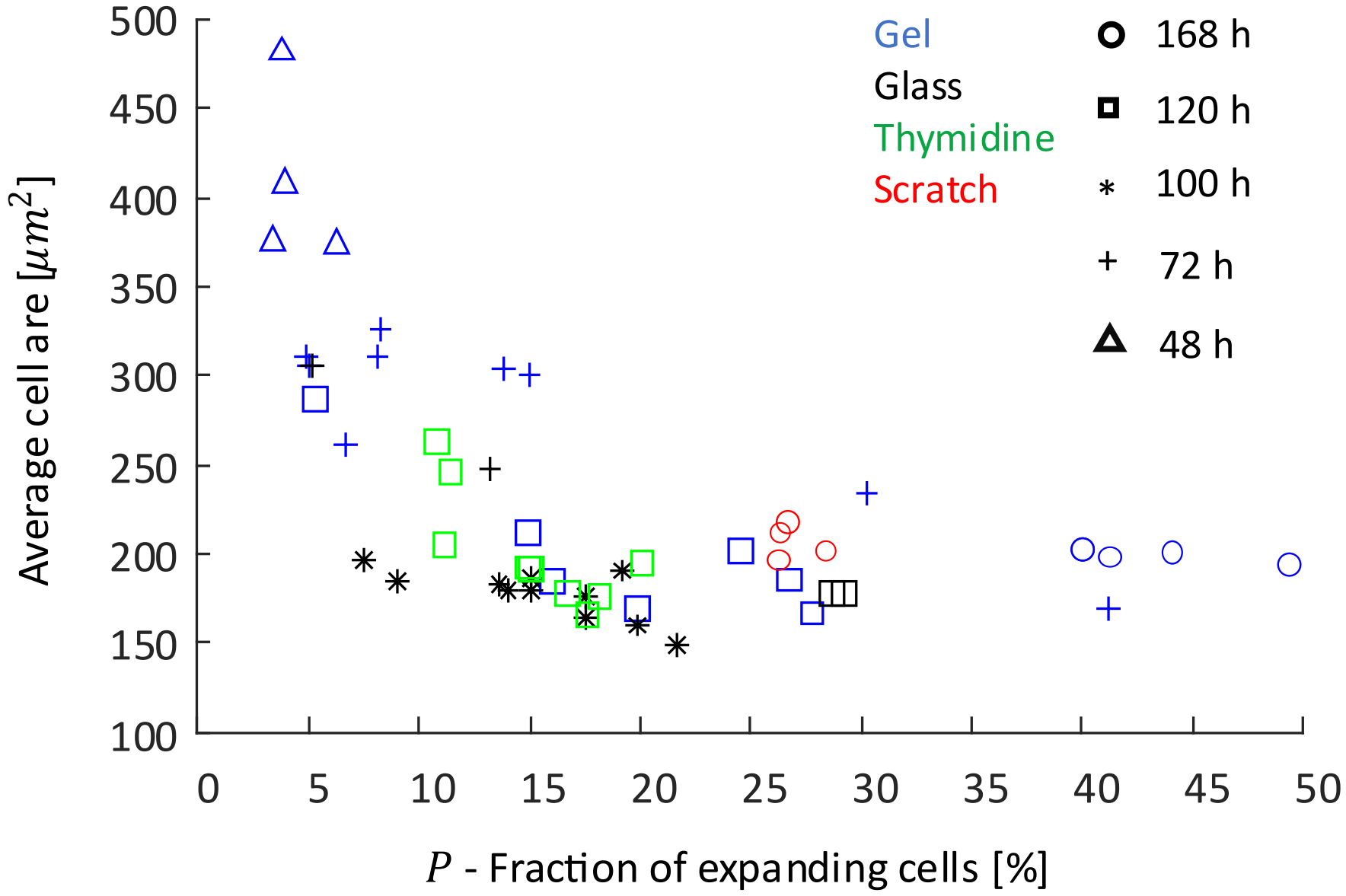

**Extended Data Fig.5 - Average cell area (in the entire tissue) as a function of expanding cells**
**population**. The data shows a continuous increase in the fraction of expending cells ($P$) in the tissue, despite
tissue density (inverse of average area) reaching a constant value as homeostasis is approached. In the
scratch experiments (red circles), the presented data is collected from outside the scratch region (Fig. 3).

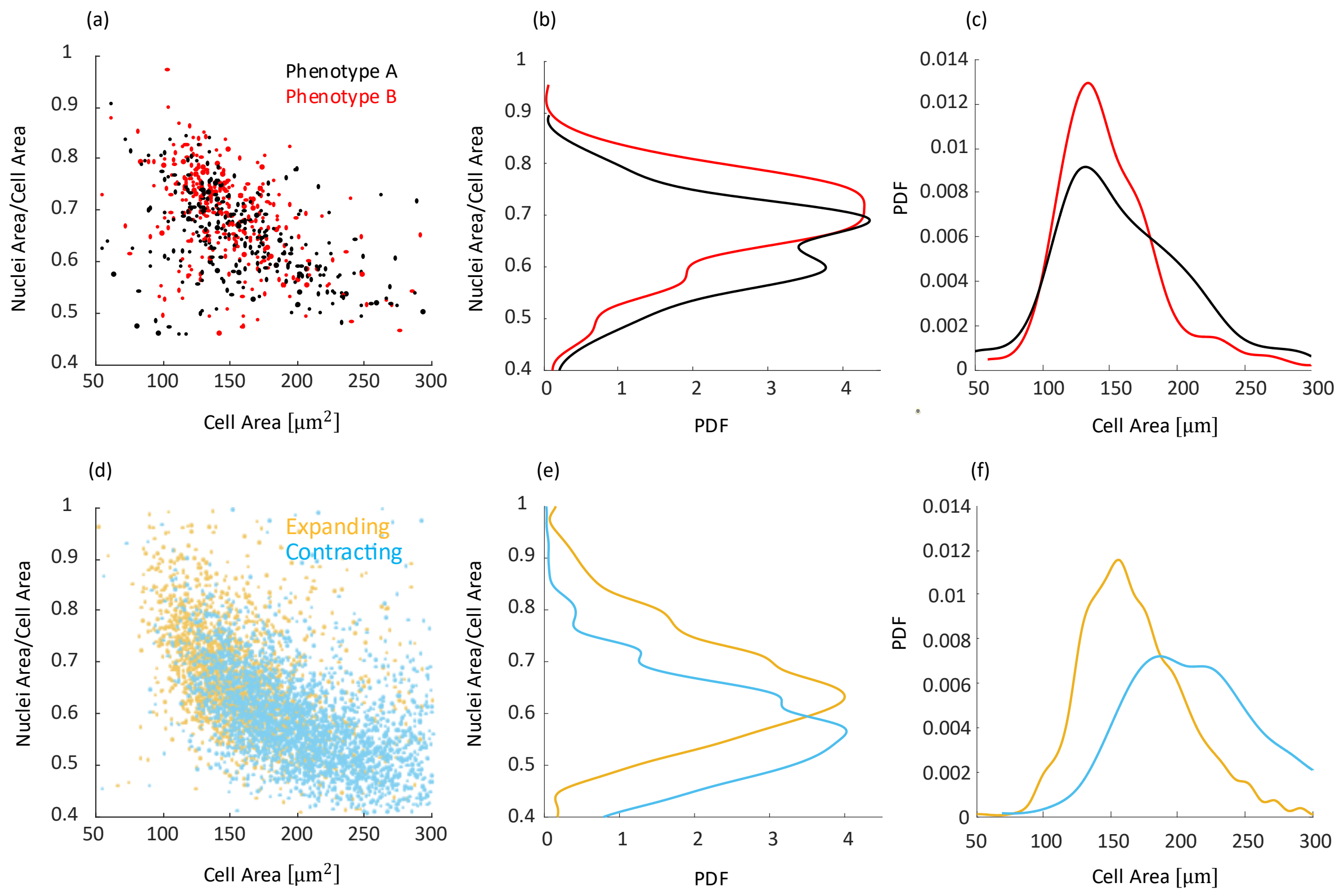

**Extended Data Fig.6 – Statistical similarity between expanding/contracting populations and B/A**
**actin phenotypes populations**. Basal-plane analysis of actin structure reveals two distinct phenotypes:
phenotype A, with actin localized at the cell cortex, and phenotype B, with actin distributed throughout the
cell. (a) Scatter plot of nucleus compactness (nucleus-to-cell-area ratio) vs cell area, and (b–c) probability
density functions (PDFs) of nucleus compactness and cell area, are shown for both A and B phenotypes.
(d–f) The same analysis is repeated for expanding and contracting cells. A and B phenotypes were clearly
distinguished by nucleus compactness (p-value<<0.00001), as well as by cell area (p-value=0.0092). Just
like A and B phenotypes, expanding and contracting phenotypes were also distinguished by nucleus
compactness (p-value<<0.00001), as well as by cell area (p-value<<0.00001). P-values are given by Mann-
Whitney Rank-sum test.

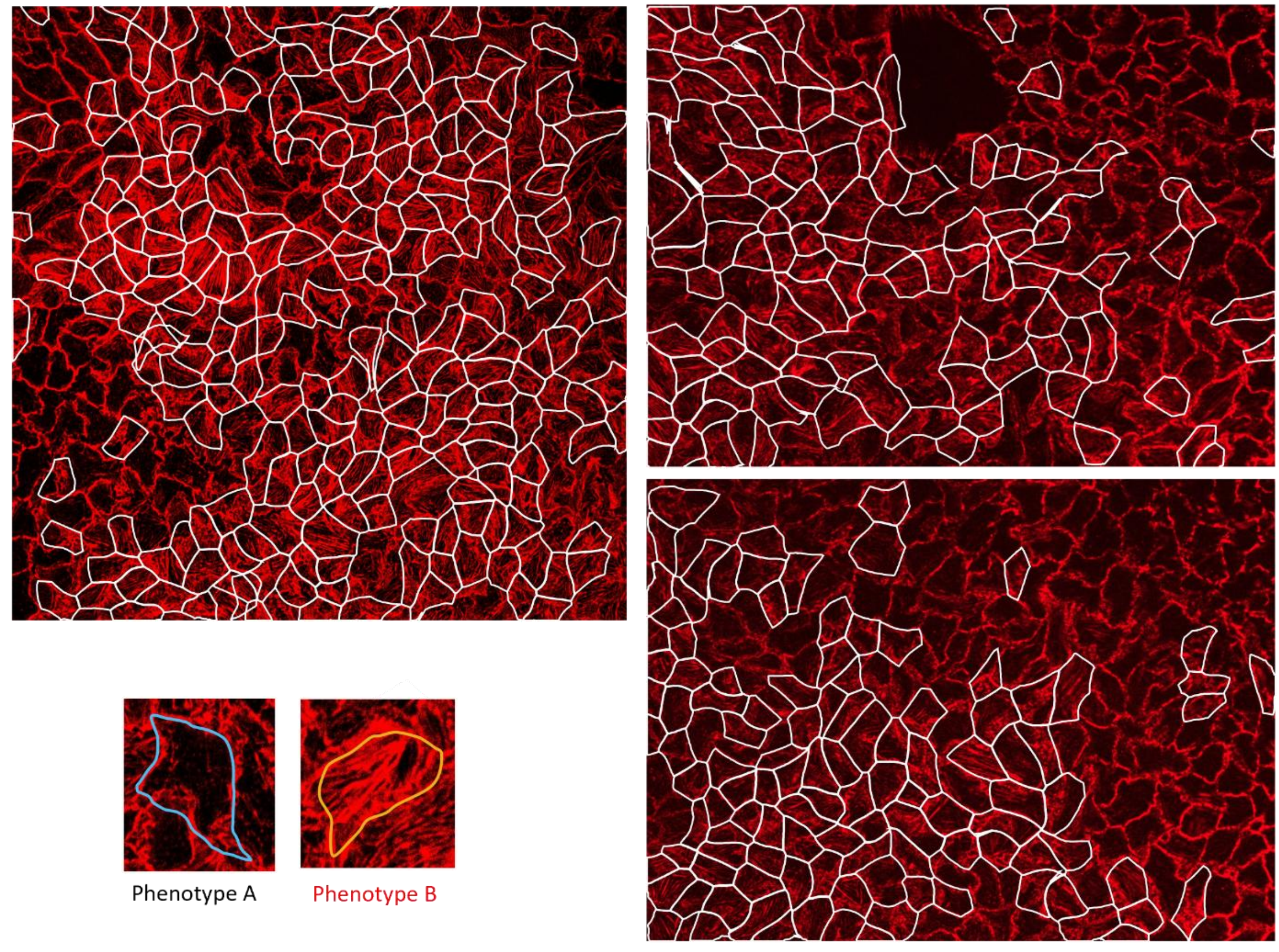

**Extended Data Fig.7 - Islands of cells with distinct actin structure.** A basal-plane view of confocal
imaging showing two distinct actin arrangements clustered as islands. White cell contours are shown for
cells that are classified as possessing phenotype B. Similar to expanding cells, phenotype B cells are
organized as clustered islands. Data is shown for three different tissues matured for 168 hours.

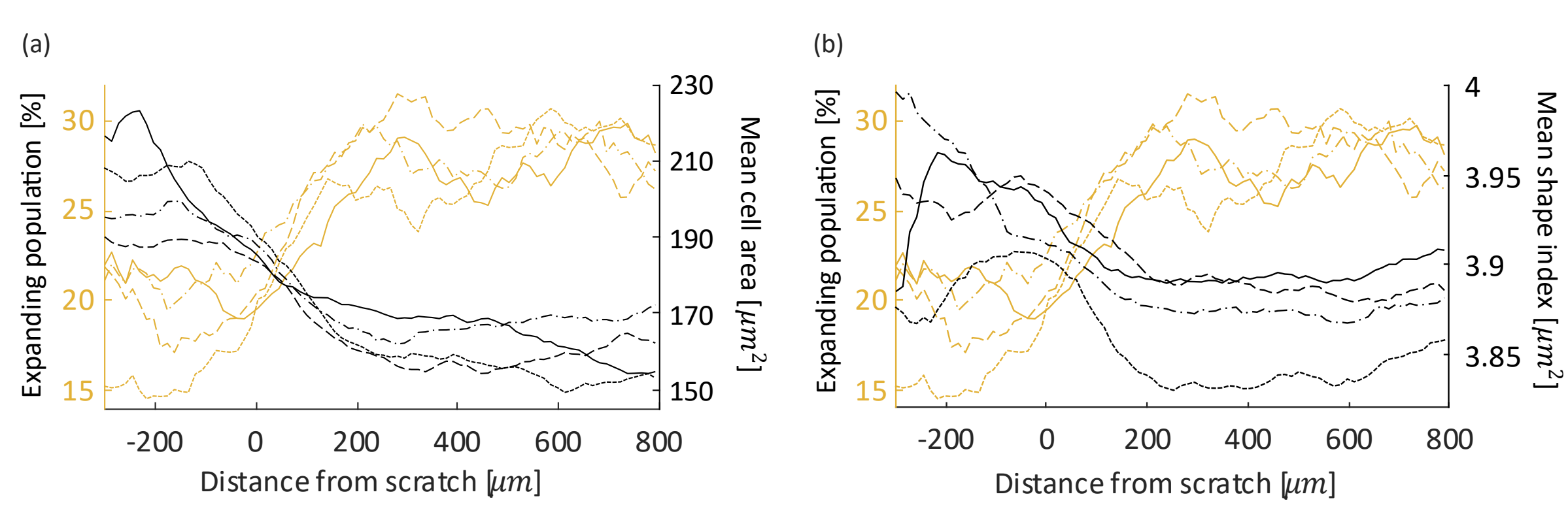

**Extended Data Fig.8 - Local jamming characteristics negatively correlate with the fraction of**
**expanding population.** During the wound healing process, as the leading edge undergoes unjamming[10-28],
the relative proportion of expanding cells (yellow curves) was averaged at different distances from the
leading edge. Expanding relative proportion is shown alongside (a) mean cell area and (b) mean shape index
(perimeter divided by square root of the area), demonstrating how jamming transition metrics negatively
correlate with the spatial distribution of expanding cells. Data were averaged within a 276 µm-wide strip at
each distance, with different curve types corresponding to different FOVs.

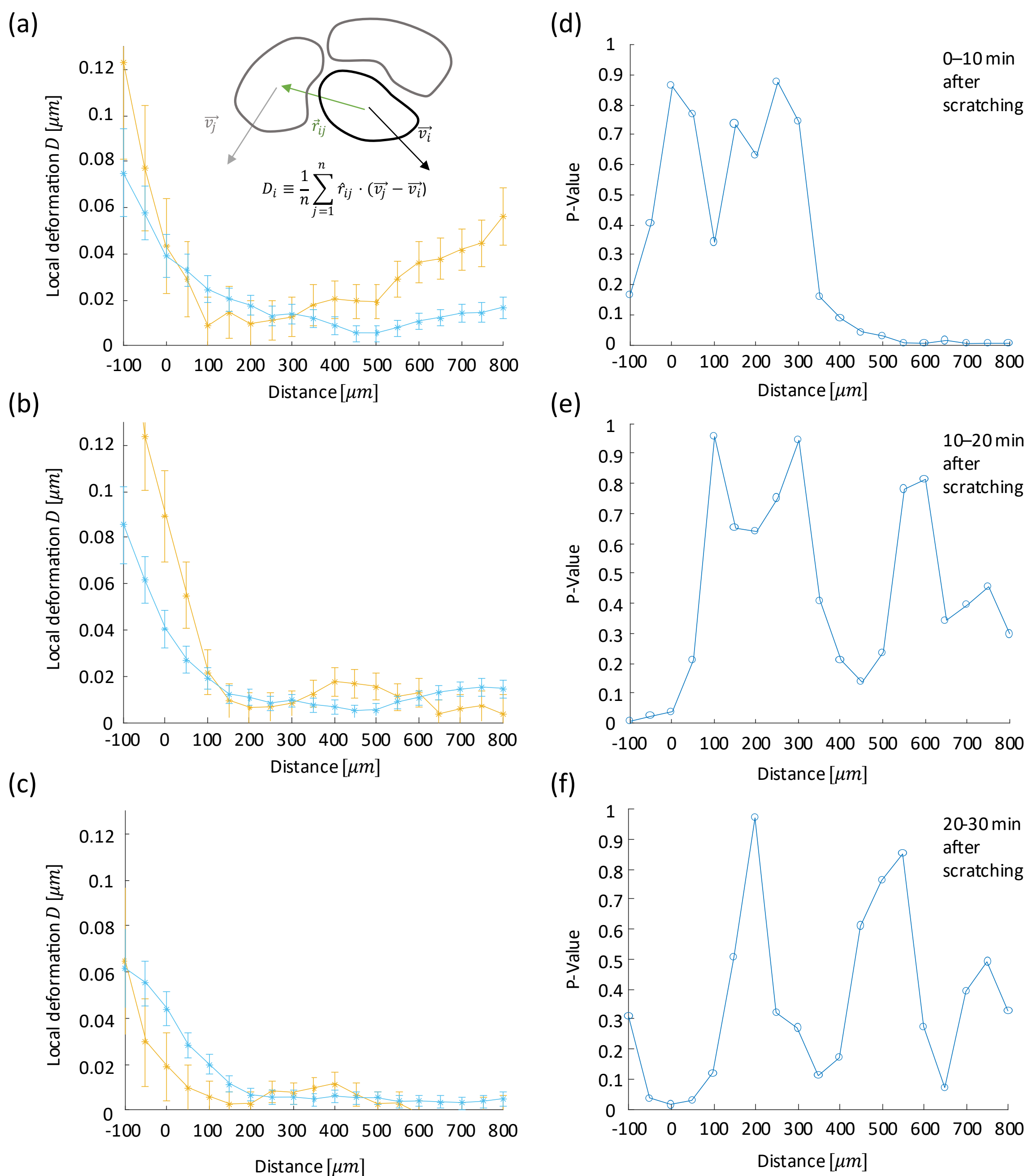

**Extended Data Fig.9 - Local tissue deformation after scratch-induced unjamming.** Deformation is
defined for each individual cell based on the relative velocities between the cell and its $n$ nearest neighbors,
projected onto the unit vector between them. A positive value indicates that neighboring cells are moving
away from each other (repulsive motion), while a negative value indicates they are moving closer (attractive
motion). These projected values were first averaged per cell, then spatially averaged according to both cell
position and mechanical phenotype (expanding or contracting). Local deformation $D$ was calculated
immediately after scratch-induced unjamming at various distances from the scratch edge. (a)–(c) show
deformation measured during the intervals of 0–10 min, 10–20 min, and 20–30 min after scratching,

respectively. (d)-(f) P-values, based on Mann-Whitney rank sum test, were calculated to assess the
significance of deformation differences between expanding and contracting cell populations. The results
indicate that differences (low P-values) are most pronounced immediately after performing the scratch (d),
particularly in regions far from the scratch edge, where expanding cells exhibit greater repulsive
deformation compared to contracting cells. This suggests that in very initial times, expanding cells that are
far away from the wound edge and are still not "mechanically informed" on the wound, exert higher inter-
tissue repulsive forces. However, as time progresses, and as the entire tissue relaxes, local tissue
deformations does not distinguish between expanding and contracting cells far from the wound, but rather
do distinguish them in proximity to the wound edge.

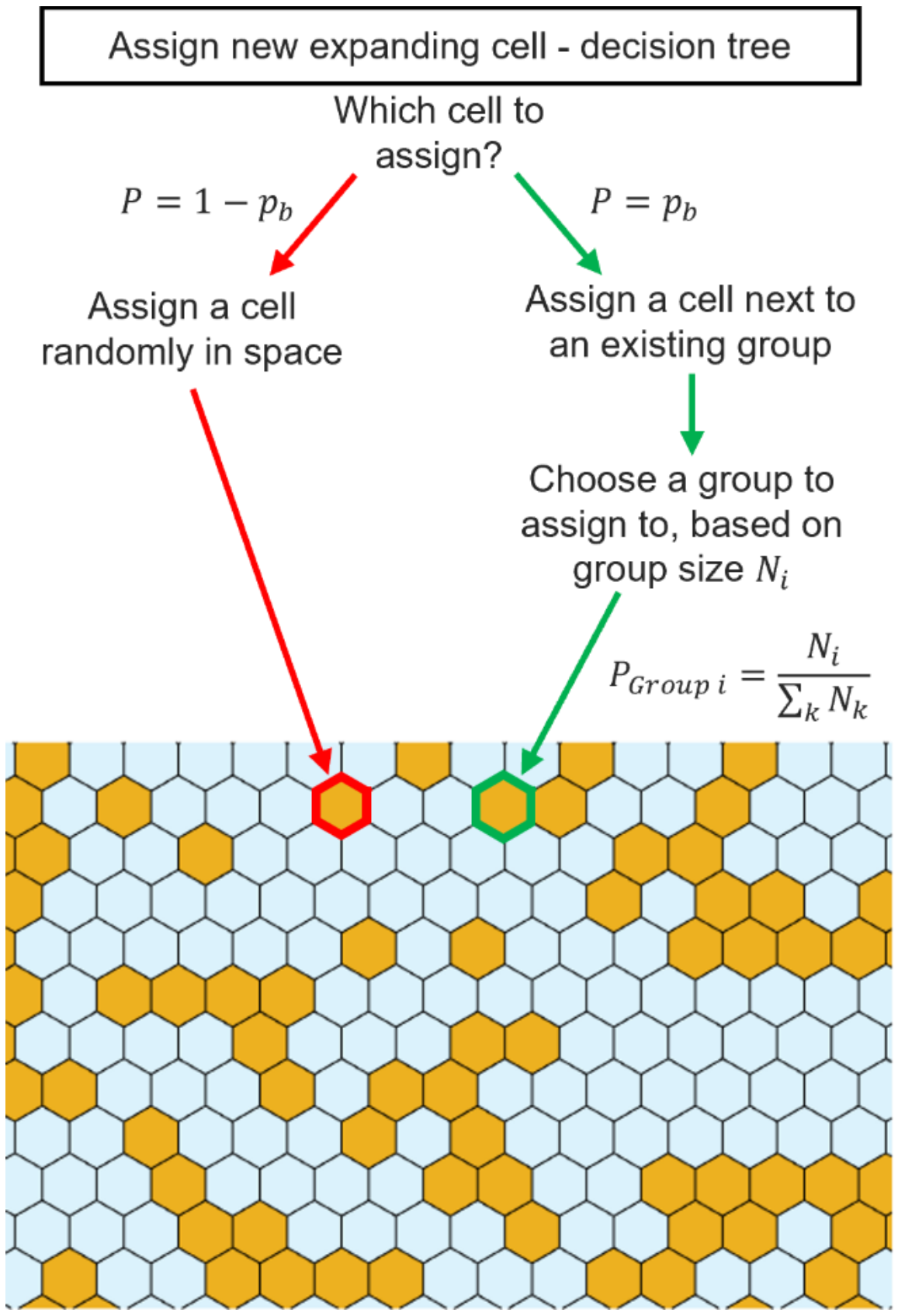

**Extended Data Fig.10 - Algorithm scheme for the site-percolation simulation.** The diagram illustrates the probabilistic assignment of expanding cells in the computer simulations, following classical site-percolation process, yet incorporating a preferential attachment bias as follows. A new "expanding cell" phenotype is placed either randomly within the hexagonal grid with probability $1 - p_b$ (red path) or preferentially attached to an existing group with probability $p_b$ (green path), based on group size. non-occupied "tensed cells" sites are shown in blue, and "expanding cells" phonotype in orange. The red and green markers highlight examples of randomly assigned and preferentially attached cells, respectively.

# Supplementary information

## A self-organized compression network arrests epithelial proliferation

Liav Daraf[1], Yael Lavi[1], Areej Saleem[1], Daniel Sevilla Sanchez[2], Yuri Feldman[1], Lior Atia[1]*

[1]Department of Mechanical Engineering, Ben-Gurion University of the Negev, Beer-Sheva 8410501, Israel;
[2]Ilse Katz Institute for Nanoscale Science & Technology, Ben-Gurion University of the Negev, Beer-Sheva 8410501, Israel;
*Corresponding author. Email: atialior@bgu.ac.il

List of supplementary figures

List of supplementary movies

Movie S1. Cells undergo deadhesion during the restrained trypsinization protocol in 120-hour matured tissue.

Movie S2. Expanding cells exhibited a higher nucleus compactness (nucleus-to-cell area ratio) than contracting cells.

Movie S3. Simulation of the network evolution incorporates preferential attachment bias.

293 # Supplementary figures

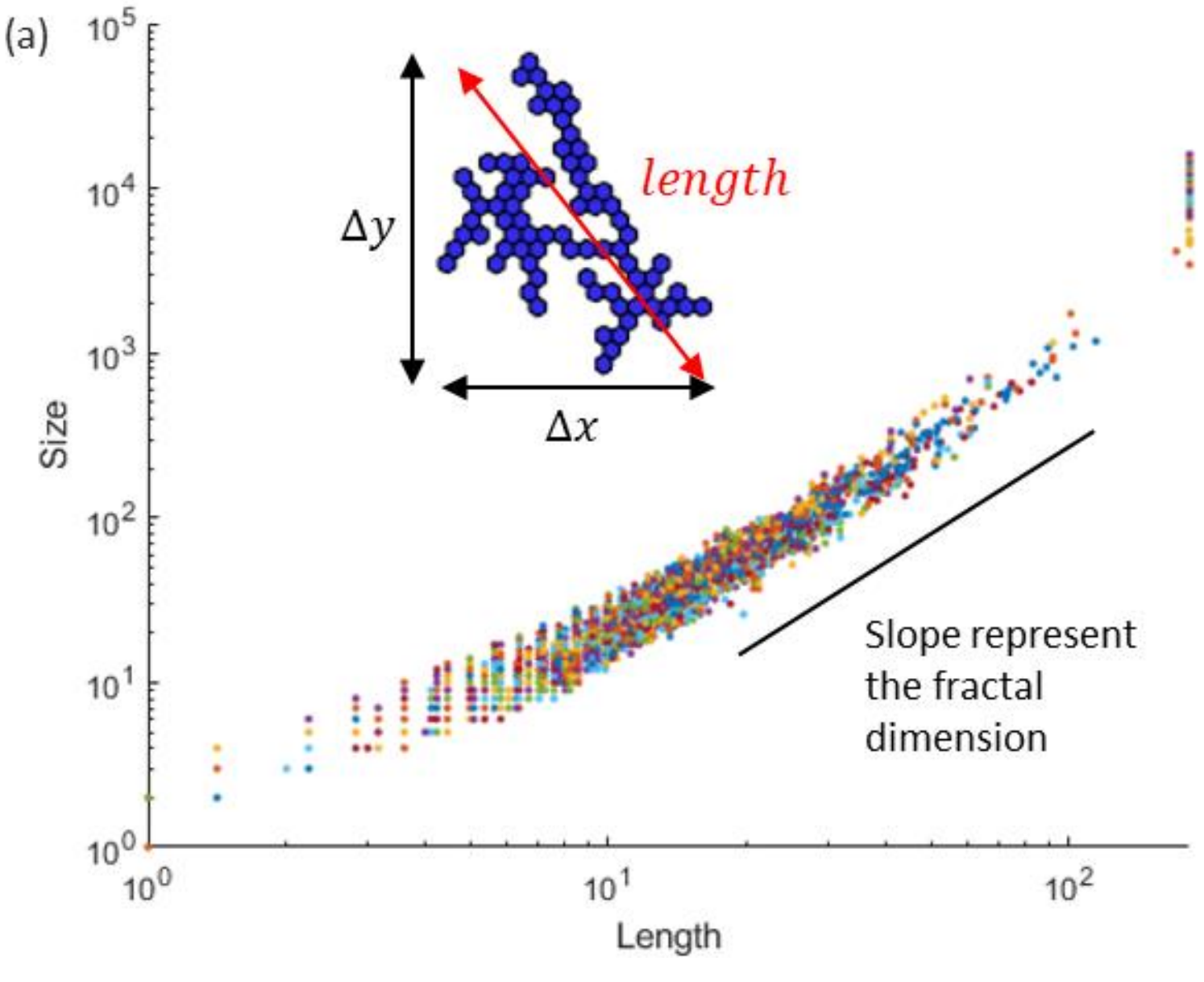

294

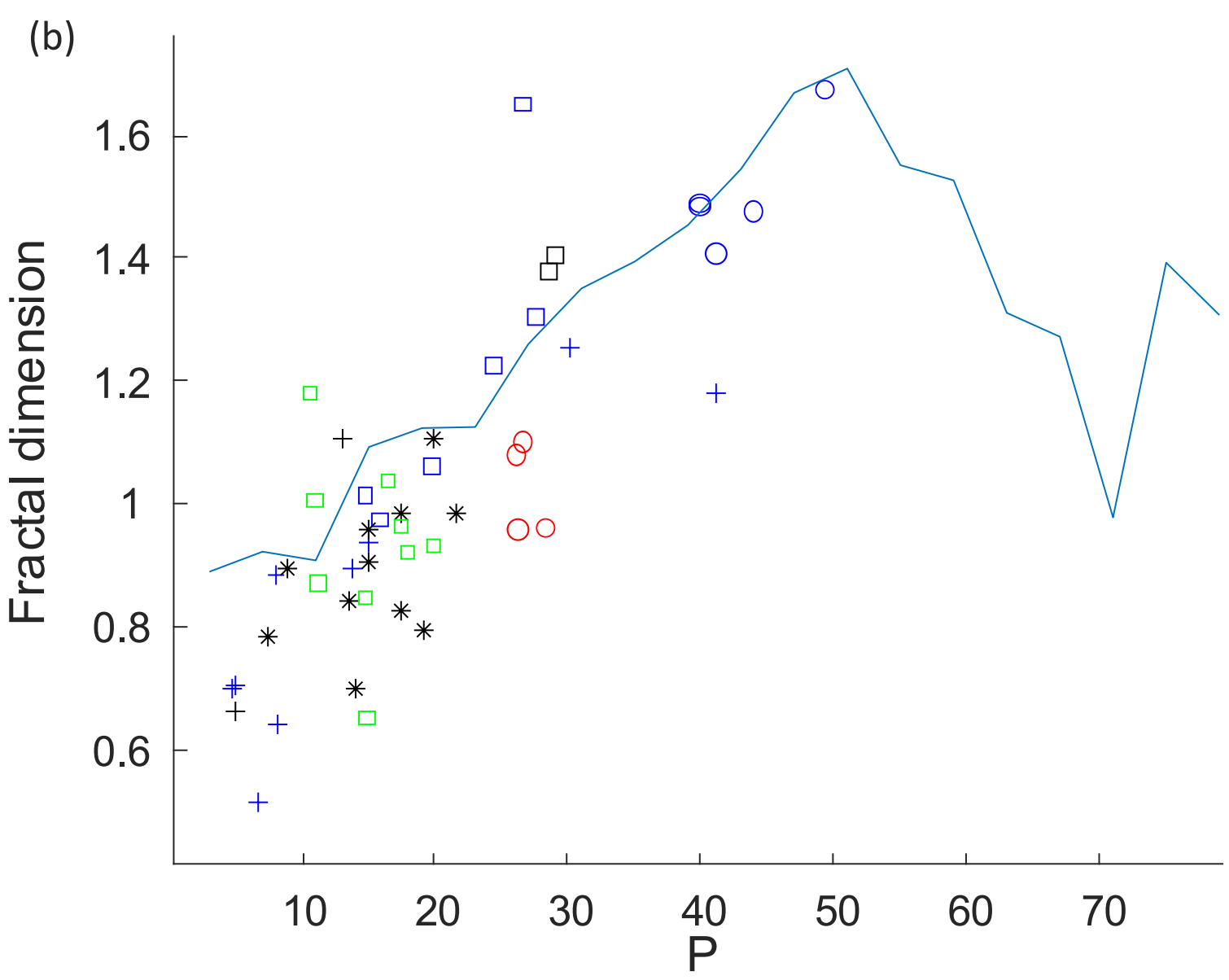

**Supplementary figure 1 - Fractal dimensions for island size.** Fractal dimension describes how the 'mass'
of an object scales with its characteristic length. The scaling typically follows a power-law relationship
$M \propto L^D$, where $D$ is the fractal dimension. In our case, the island size (or mass) refers to the number of cells
it contains, and the characteristic length is defined as the Euclidean length of its diagonal (inset). (a) Island
size is plotted as a function of island length for simulations on a 130 × 130 hexagonal grid. The fractal
dimension is estimated by the slope of the linear trend in the log-log plot, computed between the second and third quartiles of the lengths. (b) Fractal dimension as a function of occupation probability, for
experiments and simulations.

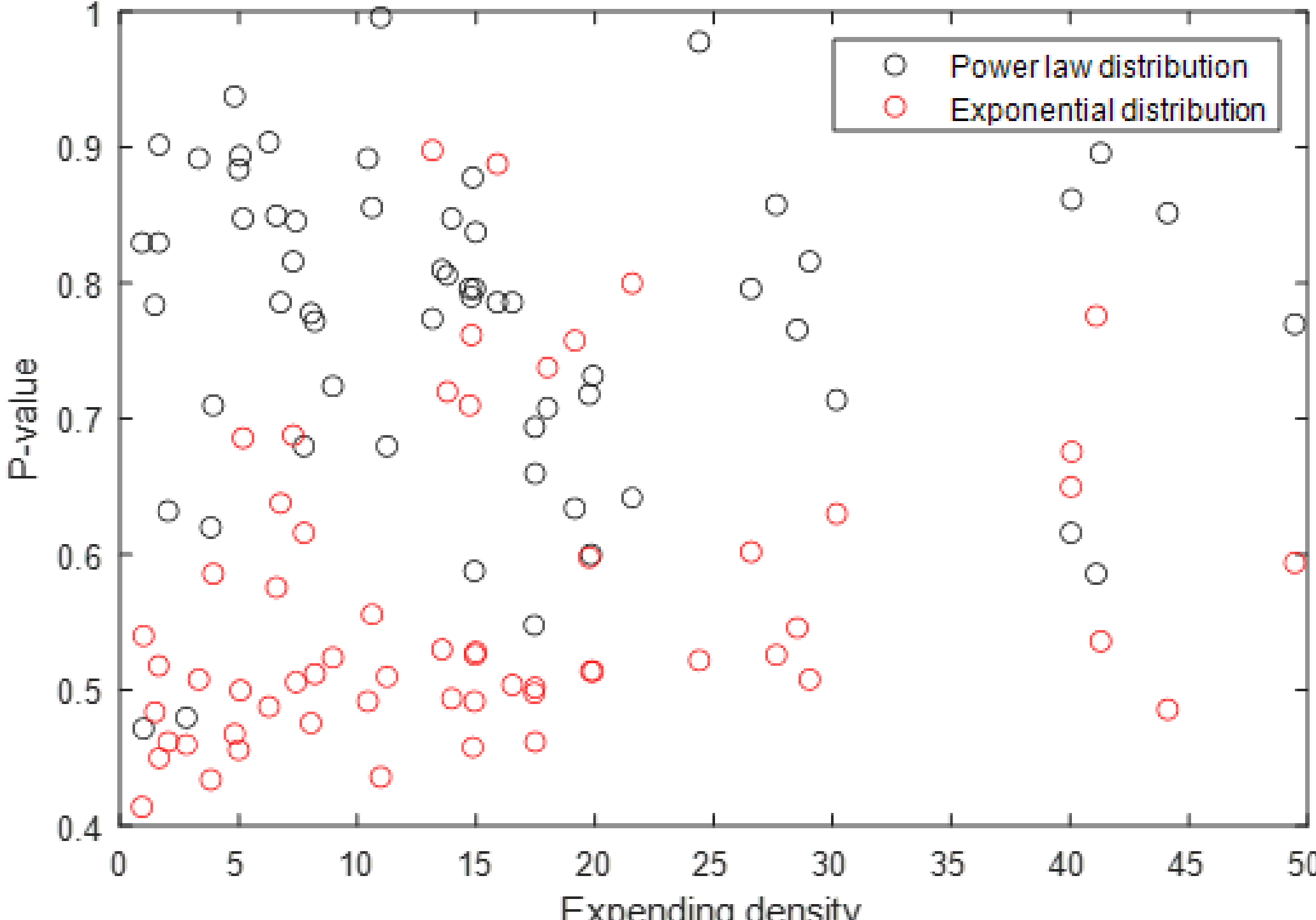

**Supplementary figure 2 - Goodness-of-fit comparison between power-law and exponential**
**distributions for group size.** The p-values for fitting group size distributions to power-law (black) and
exponential (red) forms are plotted against expanding population density (higher p-values indicate a better
fit). The results show that the power-law distribution consistently achieves substantially higher p-values,
suggesting that it provides a better description of the data.

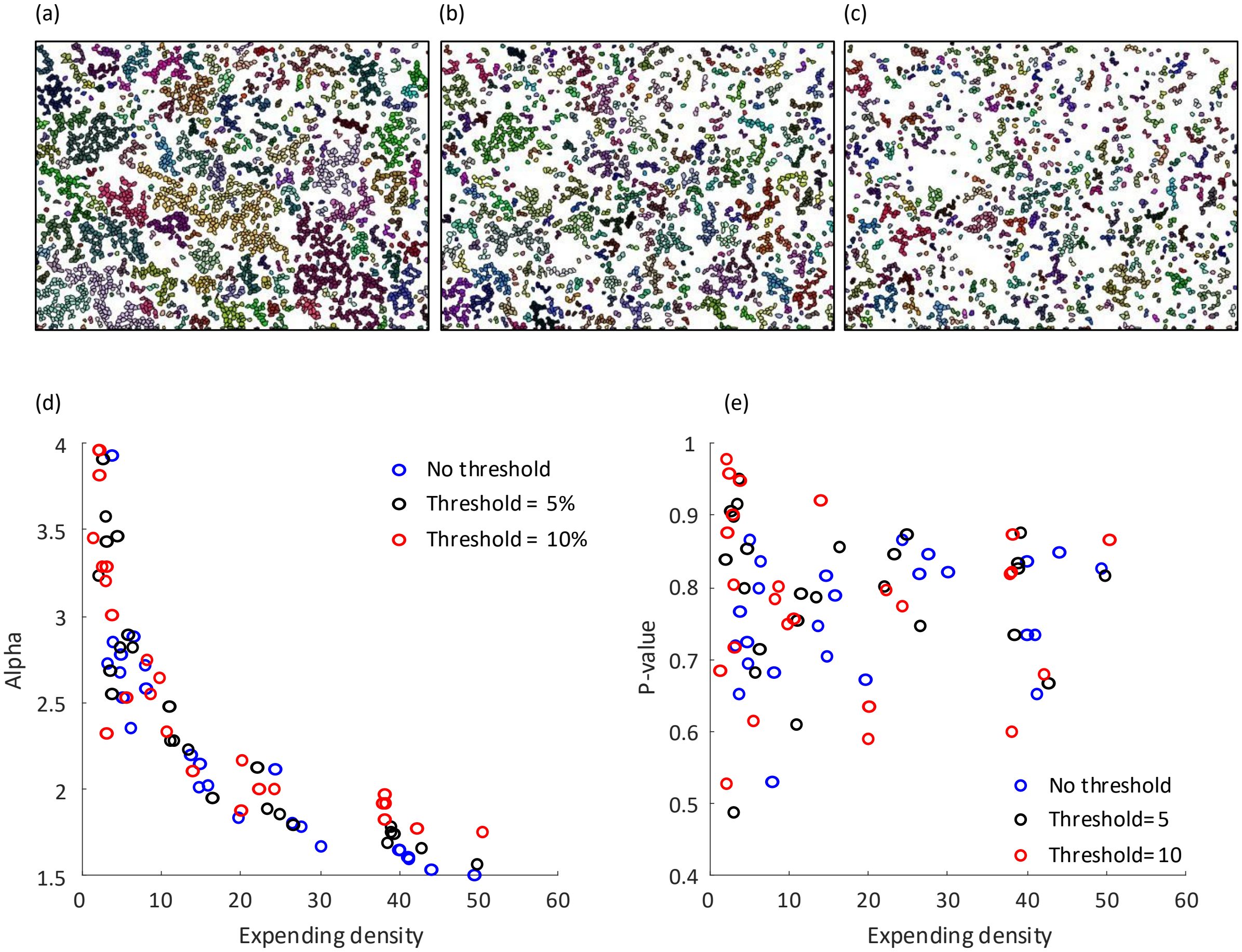

**Supplementary figure 3 - Island size distribution is insensitive to fluctuations in expanding-**
**contracting classification.** Cell classification into expanding and contracting phenotypes may be affected
by relative-area-change fluctuations around zero and measurement errors. To assess the impact of
classification uncertainty on island size distribution, we analyzed the distributions under different area-
change thresholds: (a) including all recognized cells, (b) considering only cells with an absolute area change
exceeding 5%, and (c) considering only cells with an absolute area change exceeding 10%. (d) The power-
law exponent $\alpha$ of the group size distribution is plotted against the relative proportion of the expanding
population for each threshold. (e) P-values assessing the goodness of fit to a power-law distribution are
shown as a function of expanding population density. Power-law exponent and P-value were calculated
using an iterative scheme as described in the Methods section.

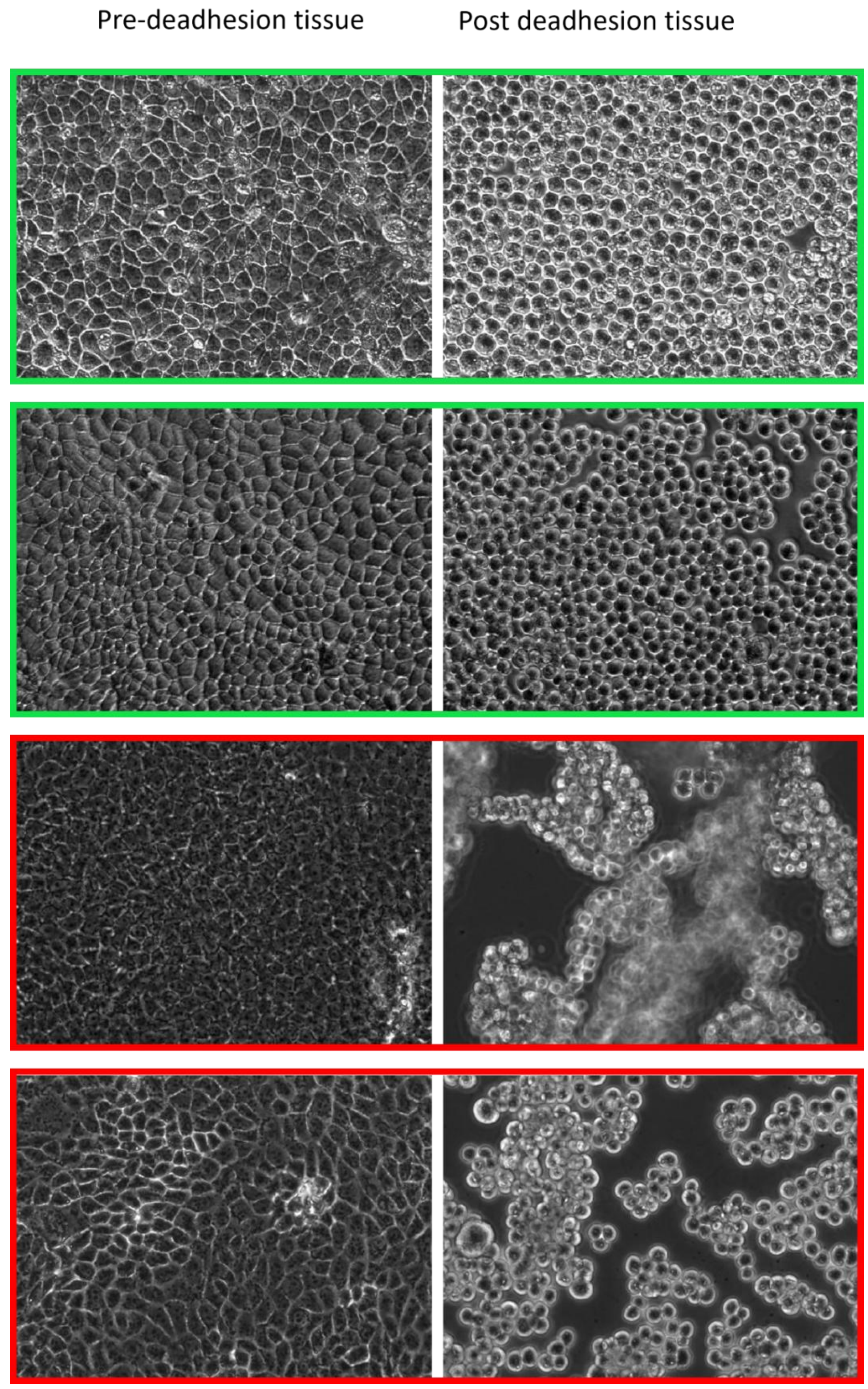

**Supplementary figure 4 - Examples of separation outcomes**: Green frames indicate satisfactory results
suitable for analysis, while red frames highlight unsatisfactory separations not suitable for analysis.

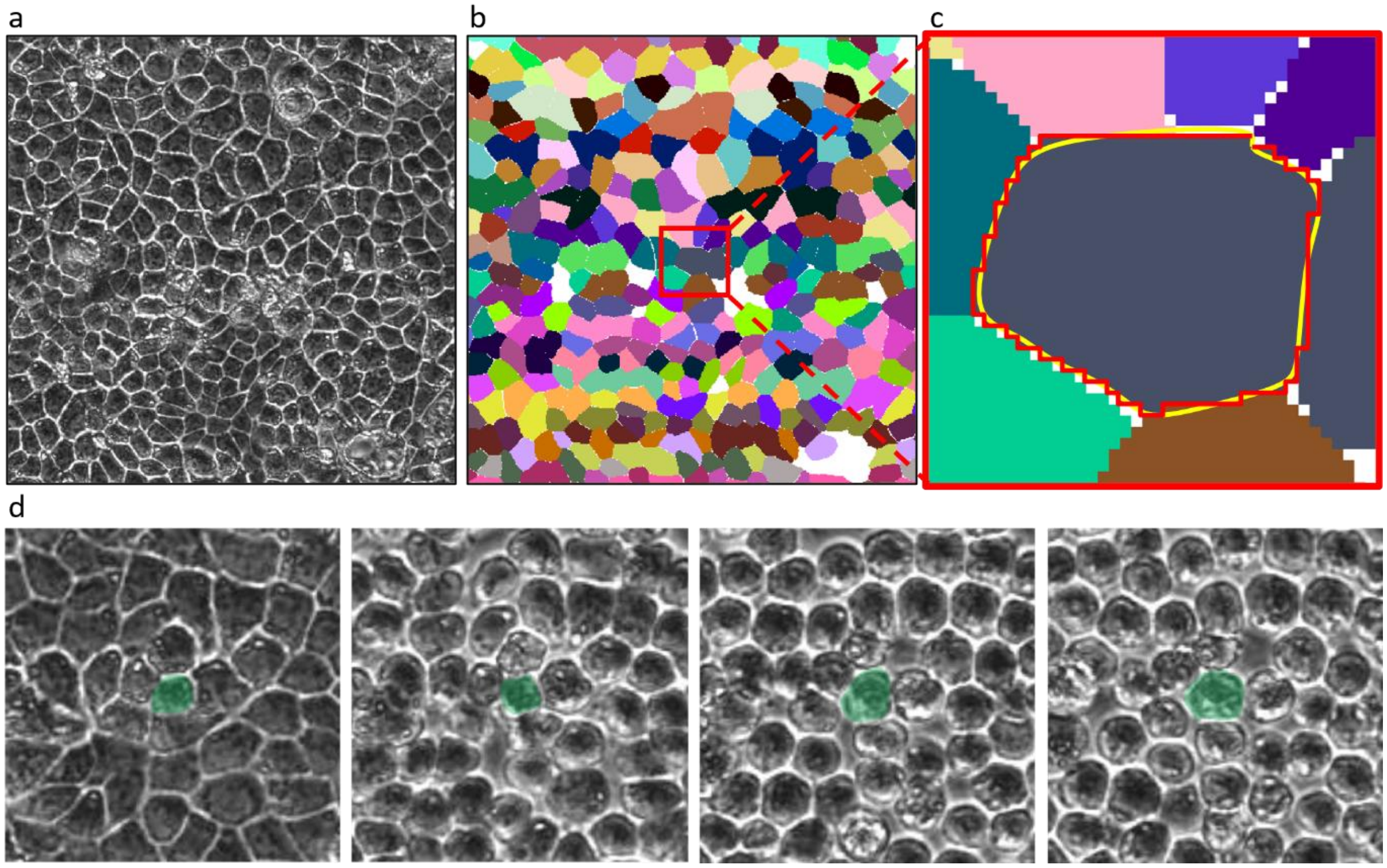

**Supplementary figure 5 - Data extraction from images.** (a) A 168-hour mature confluent monolayer as captured in our experiment. (b) Cell mask and (c) contour. The red line represents the pixelated cell outline,

and the yellow line is the spline results of the pixelated data. (d) Cell tracking across sequential frames.

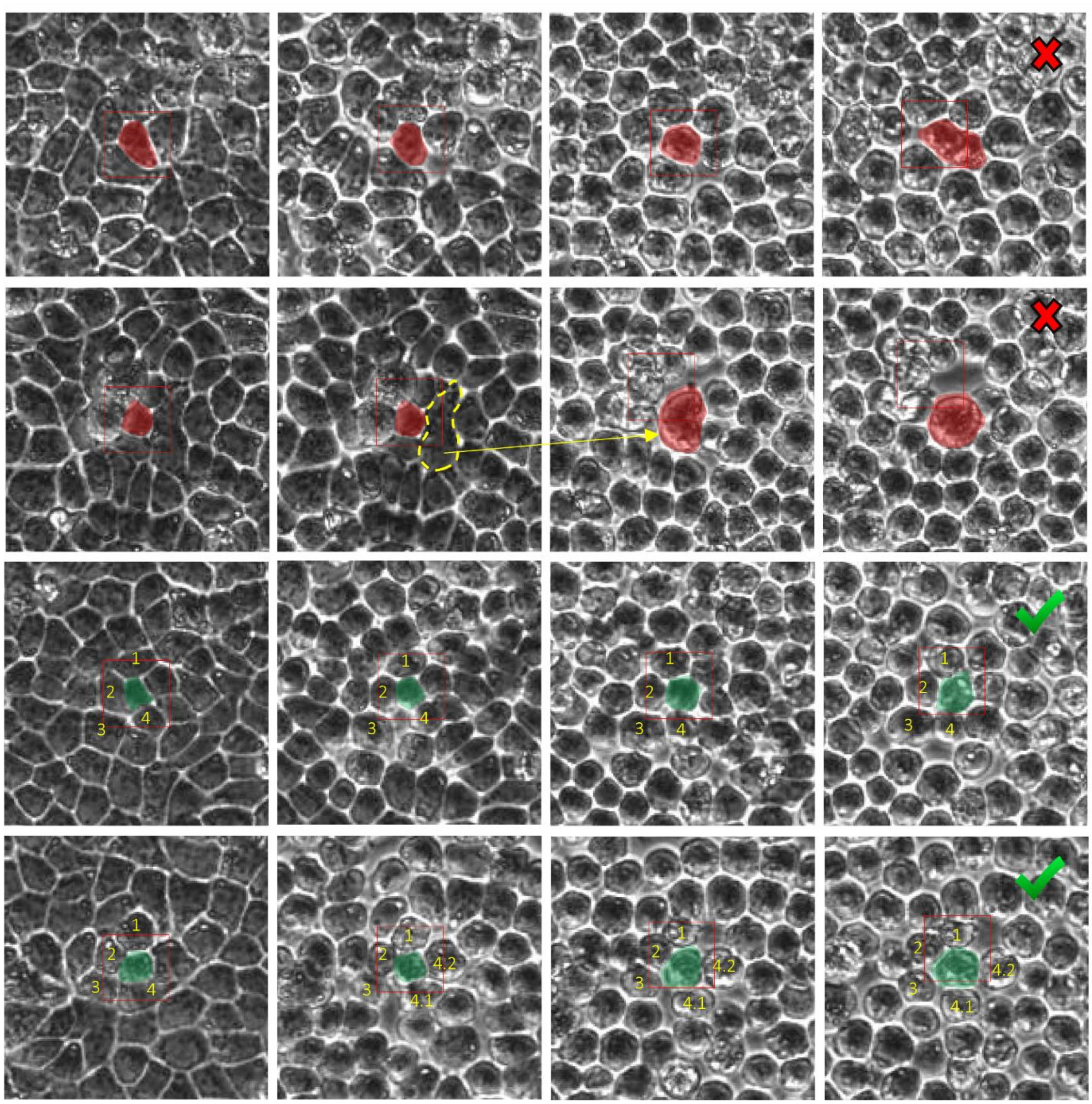

**Supplementary figure 6 - Examples of tracking examination.** Each row contains a single cell as recognized in four different time frames. For visual guide, we plotted a red frame around each examined cell's initial position. The first two rows present examples of two cells that were incorrectly identified by the tracking algorithm. In the last frame of the first row, two cells were recognized as one. In the third frame of the second row, a different cell was recognized instead of the original cell. The last two rows present examples of two cells that were successfully identified by the tracking algorithm. In both rows, we see the highlighted cell stay within the red frame which is the first visual indication of successful tracking. In addition, we can recognize the cell close neighbors (numbered in yellow) in the same orientation around the highlighted cell, which makes it possible for us to determine a successful tracking.

346

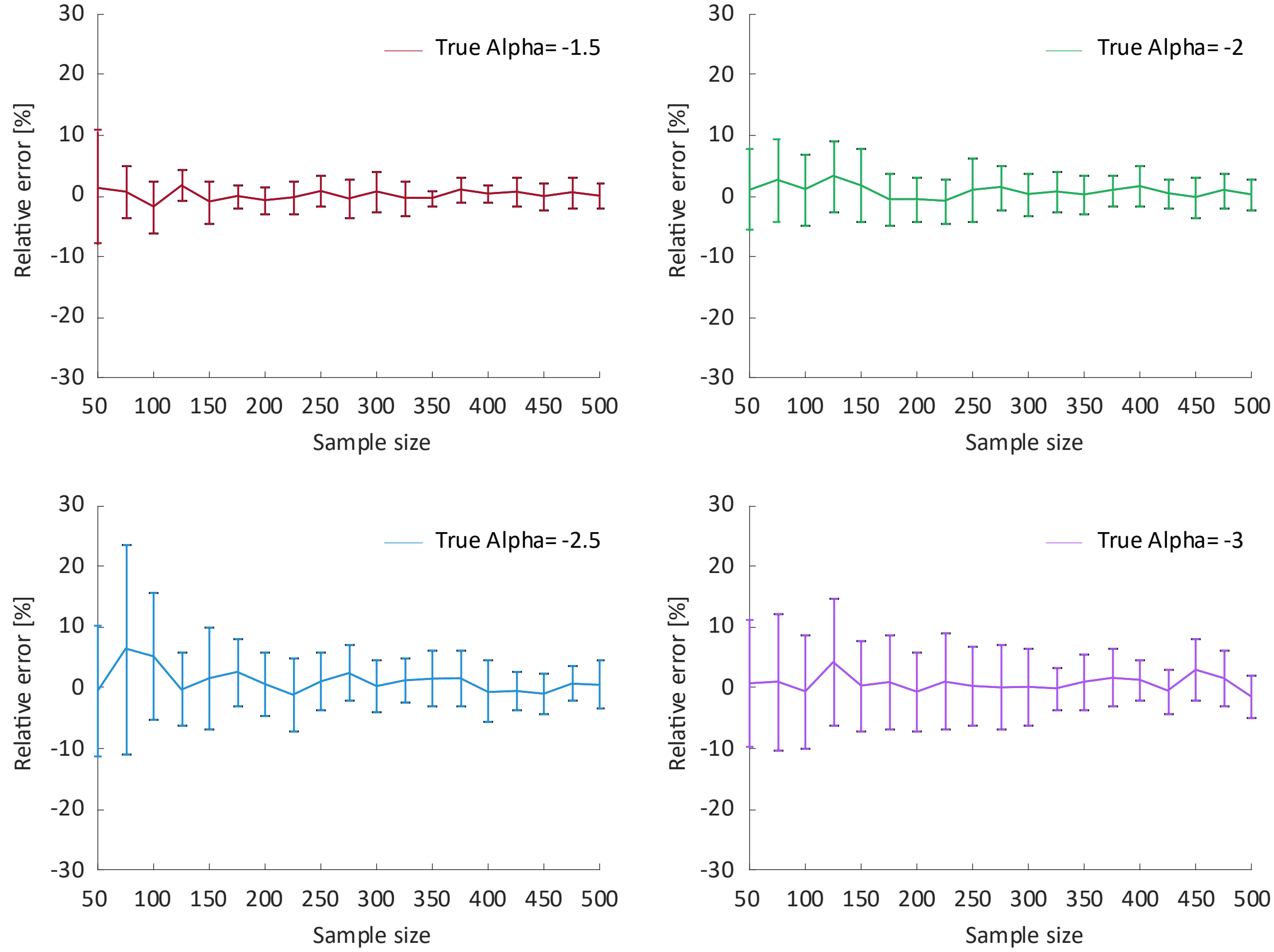

**Supplementary figure 7 - Power-law fitting accuracy across sample sizes.** Relative error in estimating
the power-law exponent (α) is shown as a function of sample size for four known synthetic distributions (α
= 1.5, 2, 2.5, 3). Data was generated using a discrete power-law with a defined cutoff and fitted using our
numerical gradient-descent scheme based on Kolmogorov-Smirnov minimization. Results demonstrate low
bias and decreasing error with increasing sample size, validating the robustness of the fitting approach for
datasets comparable in size to our experimental group size distributions. Error bars represent standard
deviation across 15 repeated simulations.

# Supplementary Movies

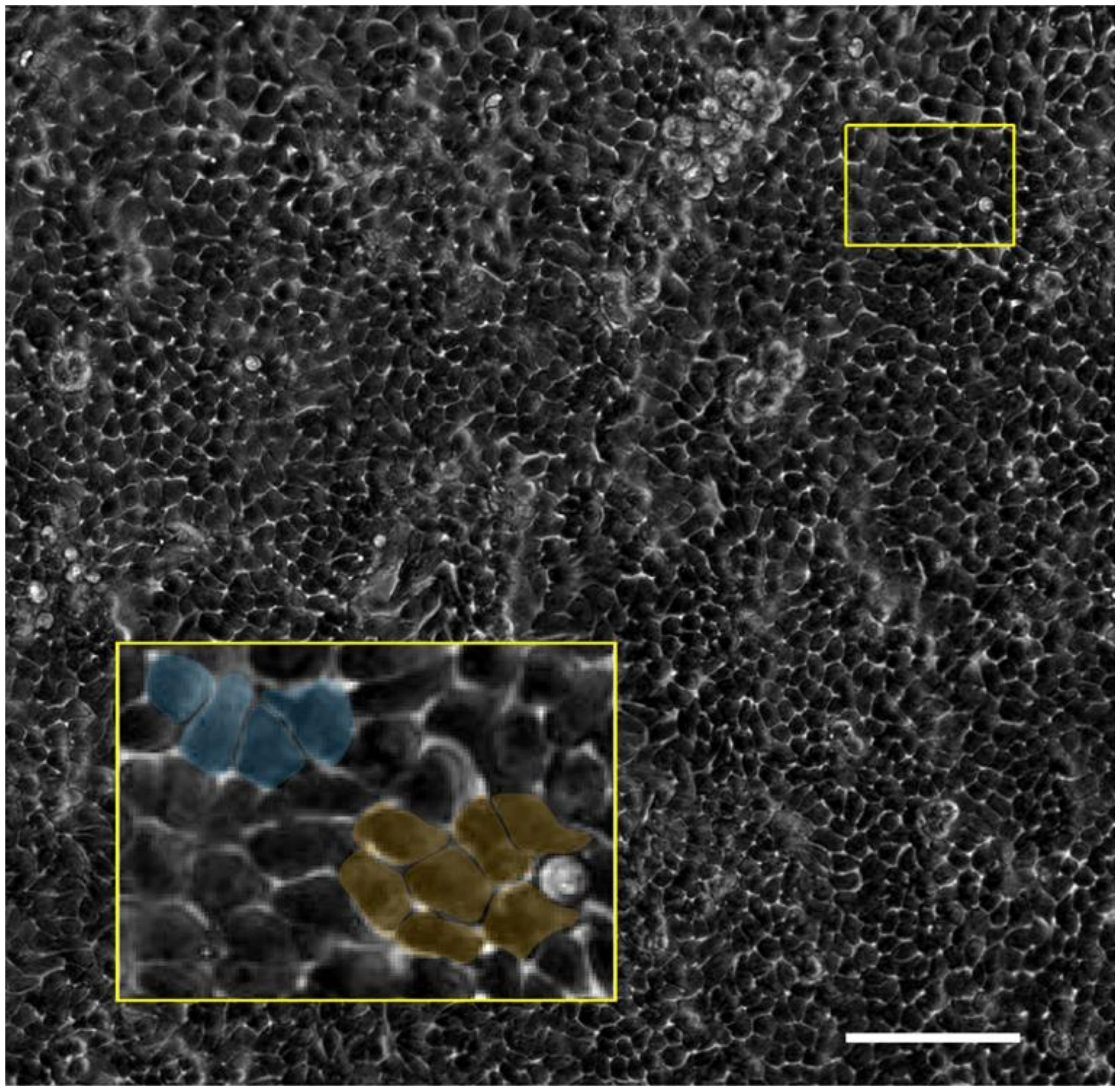

**Movie S1. Cells undergo deadhesion during the restrained trypsinization protocol in 120-hour**
**matured tissue.** In the zoomed-in region, expanding cells are marked in yellow, and contracting cells are
marked in blue. Video duration is 19 min, while the deadhesion process typically takes 15 ± 7 min (scale
bar, 100 μm).

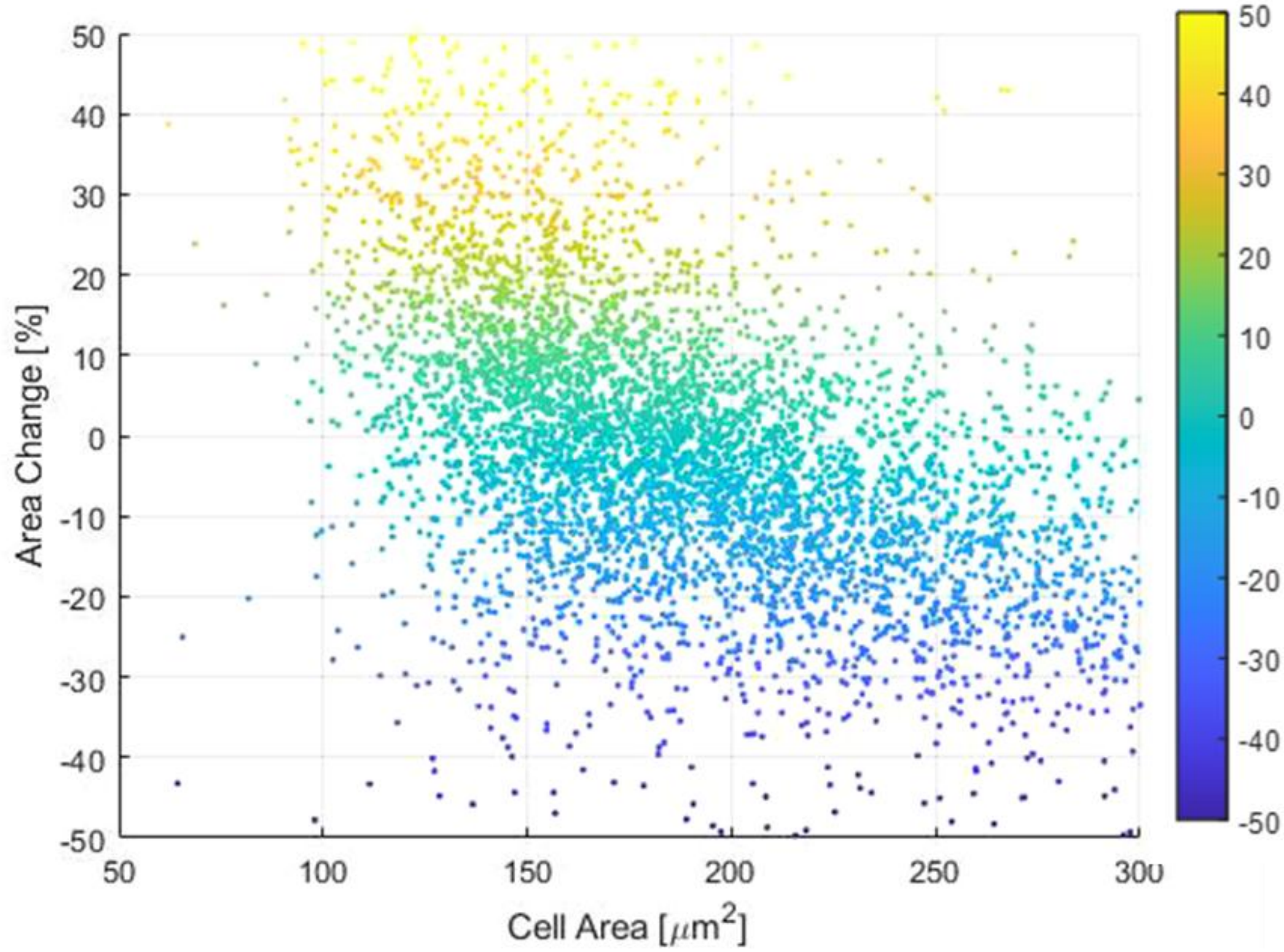

**Movie S2. Expanding cells exhibited a higher nucleus compactness (nucleus-to-cell area ratio) than**
**contracting cells.** The video initially presents a scatter plot of cell initial area versus 2D area change,
showing that small cells tend to expand while large cells tend to contract. The video then shows a third data
axis that reveals a correlation between initial cell area and nucleus-to-cell ratio, with expanding cells
exhibiting a higher compactness than contracting cells. Data from one FOV, for 168 h tissue. Colorbar
represents area change post-deadhesion.

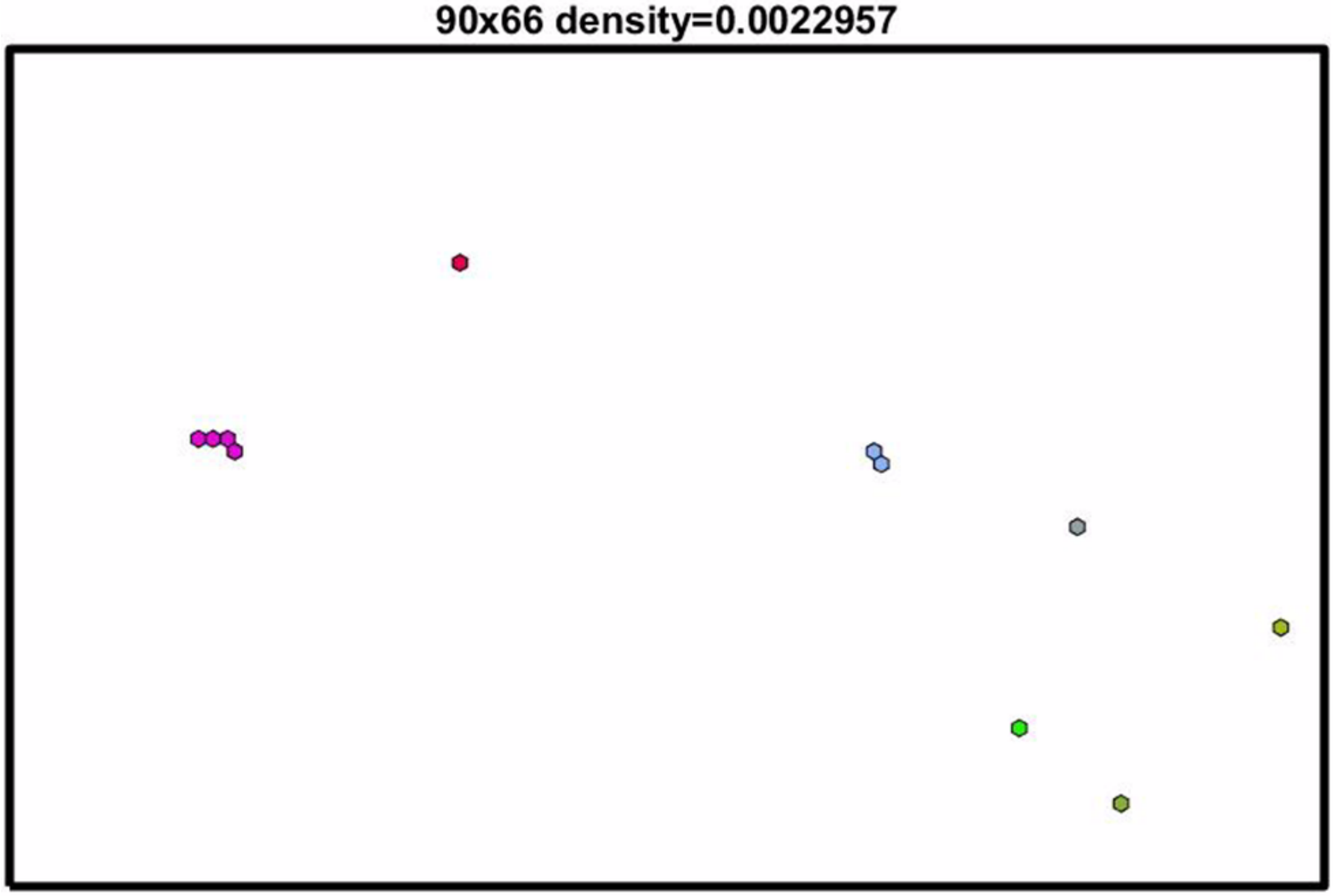

**Movie S3. Simulation of the** network **evolution incorporates preferential attachment bias.** Color-coded
groups were tracked as they expanded during the simulation. This simulation was conducted on a 90 × 66
hexagonal grid.